\documentclass[structabstract]{aa}  

\usepackage{txfonts}
\usepackage{times}
\usepackage{graphicx}
\usepackage{natbib}
\usepackage{lscape}

\def\kms{{${\rm km\;s}^{-1}$}}

\begin{document}

\title{A tale of two tails and an off-centered envelope: diffuse light around the cD galaxy NGC~3311
  in the Hydra~I cluster\thanks{Based on observations collected at the
    European Organization for Astronomical Research in the Southern
    Hemisphere, Chile, under the observing programs 082.A-0255(A),
    076.B-0641(A), 065.N-0459}}

   \author{Magda Arnaboldi\inst{1,2}
           \and Giulia Ventimiglia\inst{1,3}
           \and Enrica Iodice\inst{4} 
           \and Ortwin Gerhard\inst{3} 
           \and Lodovico Coccato\inst{1,3}}

   \offprints{M. Arnaboldi, e-mail: marnabol@eso.org}

   \institute{European Southern Observatory, Karl-Schwarzschild-Str.~2,
          D-85748 Garching, Germany
          \and INAF, Osservatorio Astronomico di Pino Torinese,
          I-10025 Pino Torinese, Italy
          \and Max-Planck-Institut f\"ur Extraterrestrische Physik,
          Postfach 1312, Giessenbachstr., D-85741 Garching, Germany
          \and INAF, Osservatorio Astronomico di Capodimonte, I-80126
          Napoli, Italy }

   \date{Accepted June~25, 2012}

\abstract
   {The formation of intracluster light and of the extended halos
     around brightest cluster galaxies is closely related to
     morphological transformation, tidal stripping, and disruption of
     galaxies in clusters.
     }
   {Here we look for observational evidence to characterize these
    processes, by studying the morphology and kinematics of the
    diffuse light in the core of the Hydra I cluster.
    }
  {We analyze Ks- and V-band surface photometry as well as deep long-slit 
  spectra, and establish a link between the structures in the light 
  distribution, the absorption line kinematics, and the LOS velocity 
  distributions of nearby galaxies and planetary nebulae (PNs).
    }
  {The central galaxy NGC 3311 is surrounded by an extended symmetric
    outer halo with $n$=10 and an additional, off-centered envelope
    at $R\sim 50''$ to the North-East. Its luminosity $L_{\rm V}= 1.2
    \times 10^{10}\, (\pm 6.0 \times 10^8)\,L_\odot$ corresponds to
    $\sim 50\%$ of the luminosity of the symmetric halo in the same
    region ($\sim 15\%$ of its entire luminosity). The velocity
    dispersion of the halo rises to cluster core values, $\sim
    400-500$ \kms, for $R>20''$. Based on measured PN velocities, at
    least part of the off-centered envelope consists of high-velocity
    accreted stars.
    
    We have also discovered two tidal streams in the cluster center,
    emerging from the dwarf galaxy HCC~026 and from the S0 galaxy
    HCC~007.  The HCC~026 stream is redshifted by $\sim 1200$ \kms\
    with respect to NGC 3311 ($V_{\rm N3311}\simeq 3800$ \kms),
    similarly as HCC~026 itself, a fraction of PNs in the off-centered
    envelope, and several other dwarf galaxies nearby. The stars in
    one of the HCC~026 tails are known to be consistent with the
    low-metallicity population of HCC~026, and our photometry shows
    that this galaxy is already mostly dissolved in the tidal field.

    The tidal stream around HCC~007 extends over at least $\sim 110$
    kpc.  It is fairly thick and is brighter on the side of the
    asymmetric outer halo of NGC 3311, which it may join. Its
    luminosity is several $10^9\,L_\odot$, similar to the luminosity
    of the stripped-down galaxy HCC~007. The redshift of the stream is
    determined from a few PN velocities and is similar to that for
    HCC~007 and HCC~026.}
  { An entire group of small galaxies is currently falling through the
    core of the Hydra I cluster and have already been partially
    dissolved by the strong tidal field. Their light is being added to
    the outer halo and intracluster light around the cD galaxy NGC
    3311. The Hydra I cluster provides a vivid example of
    morphological transformation and tidal dissolution of galaxies in
    clusters.
    }
  

   \titlerunning{Two tails and an off-centered envelope in the Hydra~I cluster core}

   \authorrunning{M.~Arnaboldi et al. }

   \maketitle

%

\section{Introduction}
Galaxy clusters are the most massive virialized structures in the
universe, and may consist of thousands of galaxies. One of the most
interesting fields in modern cosmology is the study of the
mechanisms for the growth and evolution of such systems and the
evolution of galaxies within them. The hierarchical model predicts
that structure formation and evolution occurs through the merging of
smaller units into larger systems, and this model has been supported
by many observational evidences.  The observational appearance of
clusters and their galaxies additionally depends on the evolution of
the baryonic component, which is less well understood.  As galaxies
fall into dense environments, their evolution is affected by a variety
of dynamical processes such as tidal interaction and harassment,
mergers and cannibalism, gas stripping and starvation; see
\citet{Poggianti04, DeLucia07b} for reviews. Which of these mechanisms
takes the leading role for a given galaxy morphological type in
different environmental conditions still remains to be understood.

In the nearby universe, the evolution of clusters as a whole and the
evolution of galaxies in clusters can be addressed by studying the
dynamics of the intracluster light (ICL). The ICL is the diffuse light
in galaxy clusters emitted by stars which are not bound to any
specific galaxies; for a review on the subject see \cite{AG2010}.
Wide field surface photometry shows structures in the ICL
on all scales, from a few arcminutes to degrees on the sky
\citep{tk77,Mihos05,Rudick09}. Recent studies have shown that the ICL
provides direct evidence for the dynamical status of galaxy cluster
cores \citep{Gerhard07,Doherty09,vent11}, because it contains the
fossil record of past interactions, due to its long dynamical time.

Cosmological hydro-dynamical simulations predict that the ICL is
formed by stars that are unbound from galaxies during the interactions
they experience as they fall through the cluster potential well and
interact with other cluster galaxies and the cluster tidal field. In
these simulations the ICL shows significant substructures on all
scales in its spatial and velocity distribution
\citep{Napolitano03,Murante04,Willman04,Sommer-Larsen05}.  At early
times the ICL morphology is dominated by long, linear features like
streams that become more diffuse at later times as they spread in the
cluster volume \citep{Rudick09}. Most of the simulated intracluster
stars become unbound from their parent galaxies during the merging
history leading to the formation of the brightest cluster galaxy
(BCG) in the cluster center, while for ICL at larger radii other
mechanisms like tidal stripping are more important \citep{Murante07,
  Puchwein10}.

In this paper we report surface photometry and long slit spectroscopy
of the ICL in the center of the Hydra~I cluster, a medium compact
cluster at $\sim50\;\mbox{Mpc}$ distance in the Southern hemisphere,
with a central cD galaxy, NGC 3311. The aim is to compare the
structures in the surface brightness distribution in the cluster core
around NGC 3311 with kinematic information, including also the LOS
velocities of intracluster planetary nebulas (ICPNs).  Studying the
kinematics of the ICL in nearby clusters like Hydra~I is possible with
ICPNs because (i) these objects are relatively easy to detect due to
their strong $[OIII]$ emission line \citep{Jacoby89,Ciardullo89}, and
(ii) they are good tracers of the light distribution of the parent
stellar population \citep{Buzzoni06,Coccato09}. The LOS velocity
distribution (LOSVD) of the ICPNs associated with the diffuse light in
the central 100 kpc around NGC~3311 was determined by
\cite{vent08,vent11}. They found discrete velocity components at
redshift $\sim1800$ \kms\ and $\sim5000$ \kms, in addition to a broad
component with velocity dispersion $\sigma\simeq 500$ \kms\
approximately at the systemic velocity of the Hydra~I cluster ($\simeq
3900$ \kms).  \cite{vent11} concluded that the broad velocity
component in the ICPN LOSVD may trace the high-velocity dispersion outer
halo of NGC~3311 \citep{vent10b}, while the discrete components trace
sub-components that fell through the cluster core, were tidally
disrupted, and have not yet dynamically mixed in the gravitational
potential.

This paper is structured as follows: in Section~\ref{WFIdatared} we
present optical V-band images for the Hydra~I
cluster core. Isophote fitting and analysis of the surface brightness
profiles is carried out in Section~\ref{2dphot}.  Two-dimensional
models for the optical data are derived in
Section~\ref{symmodels}, showing the existence of an off-centered
outer halo around NGC~3311 as well as tidal streams superposed on this
halo.  In Section~\ref{sec5}, long slit spectroscopic data and
kinematic measurements for the halo of NGC~3311 and the localized
stream about HCC~026 are presented and discussed.  In
Section~\ref{ICPNdwarfs} we investigate the correspondence between the
photometric components and kinematic substructures in the velocity
distribution of Hydra~I PNs, dwarf and S0 galaxies within 50~kpc of
the center of NGC~3311. Section~\ref{discussion} briefly discusses the
peculiar outer halo of NGC 3311, the properties of the newly
discovered tidal streams, and the formation of ICL from this group of
galaxies in disruption. Finally, Section~\ref{sumcon} contains our
summary and conclusions.  We assume a distance to the Hydra~I cluster
of $D=50$ Mpc, so $1'' = 0.247$ kpc.

\section{Optical imaging of the Hydra~I cluster
  core}\label{WFIdatared}

The core of the Hydra~I cluster is dominated by two giant elliptical
galaxies, NGC~3311 and NGC~3309.  Early CCD surface photometry showed
that both NGC~3311 and NGC 3309 are fitted by an R$^{1/4}$ law within
$30''$ distance from their centers \citep{vast91}. On the basis of the
large $R_e = 98''$ value from the R$^{1/4}$ fit, \cite{vast91}
classified NGC~3311 as a cD galaxy.  In the center of this galaxy, a
complex dust lane is found, and the observed surface brightness
profile thus falls below the best fit R$^{1/4}$ profile. Both the dust
lane and the core in the central profile are confirmed by HST WFPC2
imaging \citep{laine+03}. NGC~3309 does not have a dust lane in its
core \citep{vast91}.  \cite{Misgeld08} published VLT/FORS1 images of
the Hydra cluster in the V-band, that were used by \cite{Richtler11}
to derive the average radial surface brightness profile for NGC~3311
down to $\mu_V= 26.5$ mag arcsec$^{-2}$.  They classified NGC~3311 as
a cD galaxy because of the low central surface brightness and extended
radial profile.

In the present work we derive accurate quantitative photometry for the
galaxy NGC~3311, NGC~3309 and the diffuse light in the Hydra~I core.
In what follows, we use two data sets for this purpose: VLT/FORS1
archive data for the center of the Hydra~I cluster and V band images
obtained with WFI at the ESO/MPI 2.2m telescope.  These V band images
allow us to measure the profile shape at large radii where the surface
brightness is low, and analyze the substructures in the light
distribution.  The VLT/FORS1 and 2.2m/WFI data contain complementary
information: the FORS1 data are deep, and allow reliable measurements
down to $\mu_V = 26.5\, \mbox{mag}\, \mbox{arcsec}^{-2}$
\citep{Richtler11}, while the WFI data with their large field-of-view
provide a robust estimate of the sky background.

\begin{figure*}[hbt!] \centering
\includegraphics[width=16.5cm]{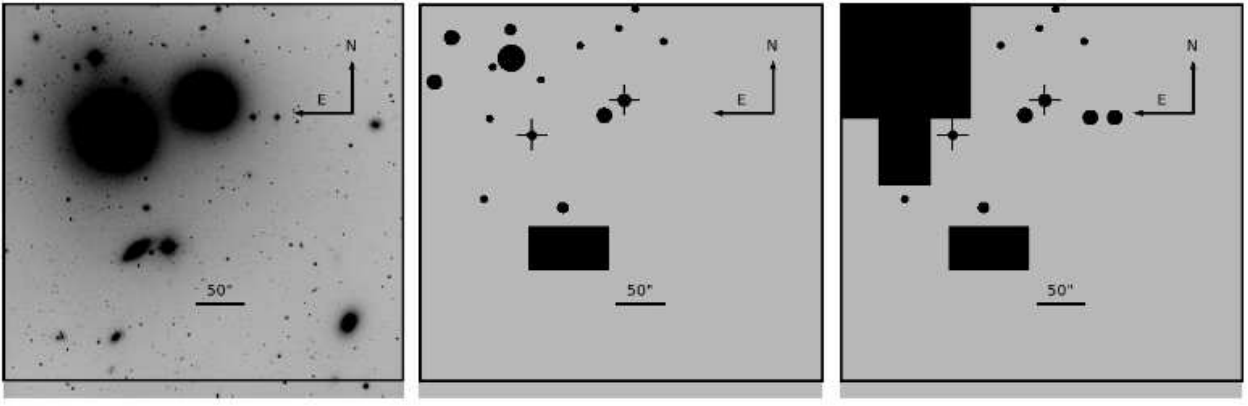}
\caption{ Left panel - VLT/FORS1 V band co-added image of the Hydra~I
  cluster core with NGC~3311 and NGC~3309 visible in the upper part of
  the image. A 2D-photometric model is fit within the region of $6'.8 \times
  6'.4$ limited by the box (see Sect. ~\ref{2dgalfitv} for details).
  Middle panel - The mask ``cmask'' adopted for the 2D-model fit of the
  VLT/FORS1 data, which masks stars and galaxies in the chosen
  region as well as the centers of NGC~3311 and NGC~3309.  Right panel
  - The mask ``allmask'' for the 2D fit that masks also part of the
  North-East regions in addition to the stars and galaxies in
  ``cmask'' (see Sect.\ref{2dgalfitv} for
  details). In the middle and right panels, the two crosses indicate
  the positions of the centers of the two giant galaxies.}
\label{VFORS1}
\end{figure*}

\subsection{VLT/FORS1 deep V band photometry - Observations and 
data  reduction}

Johnson V band imaging data for NGC~3311 and NGC~3309 were retrieved
from the ESO Science Archive Facility, acquired with VLT/FORS1 in
service mode on April 04, 2000 (program 65.N-0459(A), PI Hilker).  The
VLT/FORS1 images cover a field-of-view (FoV) of $6.8'\times6.8'$ at an
angular scale of $0''.2$ pixel$^{-1}$. During the observations, the
FoV was centered at $\alpha=10\mbox{h}36\mbox{m}36.14\mbox{s},\;
\delta=-27\mbox{d}32\mbox{m}51\mbox{s}$, avoiding a bright star NE of
NGC~3311. The dataset included three exposures of 480 sec each for a
total observing time of $0.4$ hr. The resulting average seeing in the
combined median image is $FWHM \sim 0''.6$; the central $\sim 5''$
diameter region of NGC~3311 and $\sim 8''$ diameter region of NGC~3309
are saturated.

Standard calibrations, bias, and sky flats were also retrieved from
the ESO Archive.  Several Landolt standard stars in the Rubin 152
field were observed in V band for the photometric calibration. The
zero point for the V band VLT/FORS1 photometry $ZP_{V,FORS1}=27.43\pm
0.06$ is derived independently in the present work. The correction for
extinction in the V band amounts to $0.25$ mags \citep{sch+98}.

The instrumental signature of the VLT/FORS1 images was removed using
standard IRAF tasks. Spikes from saturated stars in the field were also
removed; see Section~\ref{Vband}. Images were registered before
co-addition; the co-added final VLT/FORS1 V band image is shown in
Fig.~\ref{VFORS1}. 

\subsection{Wide Field Imager V band photometry - Observations and
  data reduction}\label{Vband}

Johnson V band imaging of the Hydra~I cluster was acquired
in service mode on the night of January 12, 2006 at the Wide Field
Imager (WFI) on the ESO/MPI 2.2m telescope, at the La Silla
observatory. The WFI imager covers a field-of-view (FoV) of
$34'\times33'$ with a mosaic of 4$\times$2 CCDs (2k$\times$4k) at an
angular scale of $0''.238$ pixel$^{-1}$. During the observations, the
FoV was centered at $\alpha=10\mbox{h}36\mbox{m}51\mbox{s},\;
\delta=-27\mbox{d}31\mbox{m}35\mbox{s}$. 13 exposures of 300 sec each
were obtained for a total observing time of $\sim 1.1$ hr. The
resulting average seeing in the combined median image is $FWHM \sim
0''.7$.

Standard calibrations, bias, sky flats and dark skies, were also
obtained.  Several Landolt standard stars in the Rubin 149 field were
observed in V band for the photometric calibration. The zero point for 
the V band photometry is $ZP_V=24.02\pm 0.02$.

\begin{figure*}[hbt!] \centering
\includegraphics[width=16.5cm]{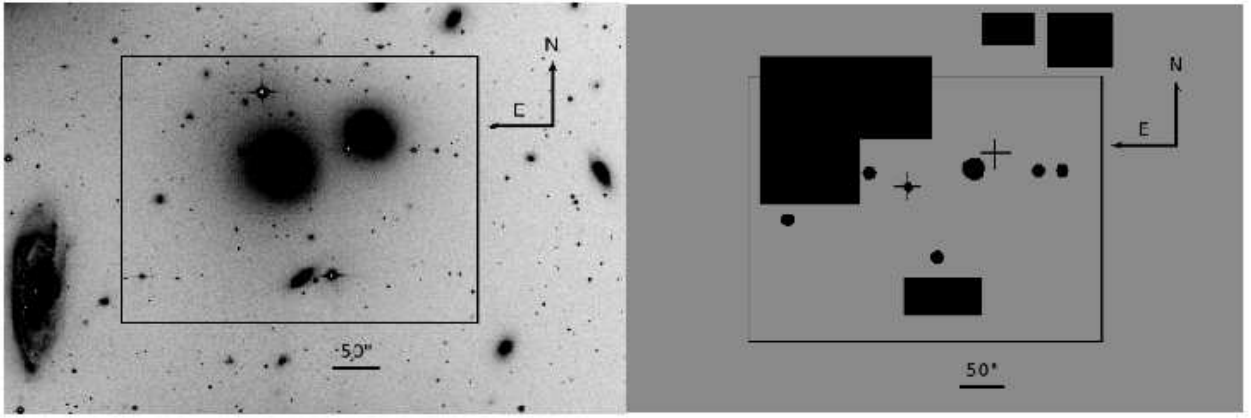}
\caption{ Left panel - 2.2m/WFI V band image of the Hydra~I cluster
  core; NGC~3311 is visible on the left and NGC~3309 on the right. A
  2D-photometric model is fit within the region of $6'.4 \times 4'.8$
  limited by the box.  Right panel - The mask ``allmask'' adopted for
  the fit to the bright central regions (see Sect.~\ref{1dplus2d} for
  details). The two crosses in the right panel indicate the positions
  of the centers of NGC~3311 and NGC~3309.}
\label{Vfinal}
\end{figure*}

Data reduction is carried out with standard IRAF tasks for
pre-reduction and calibration. After bias subtraction and flat
fielding, the frames are corrected for any residuals by using dark
skies.  Then the average background emission is measured in several
regions of the FoV far from the galaxy light and the final average
value is subtracted off each single frame. The IRAF task {\it
  NOAO.NPROTO.IRMOSAIC} is used to obtain a mosaic of the CCDs.  Each
frame is inspected and bright spikes associated with saturated stars
are removed by linear interpolation from nearby columns/rows before
frames are co-added. Residual sky offsets with respect to the average
value are removed before the CCD frames are mosaiced.  Finally, the
mosaic images are registered. Then image fluxes are scaled to a
reference image to account for transparency variations, following
which images are co-added with a threshold rejection implemented to
reduce effects from CCD gaps.  The final co-added image is shown in the
left panel of Fig.~\ref{Vfinal}.

\begin{figure}[hbt!] \centering
\includegraphics[width=8.5cm]{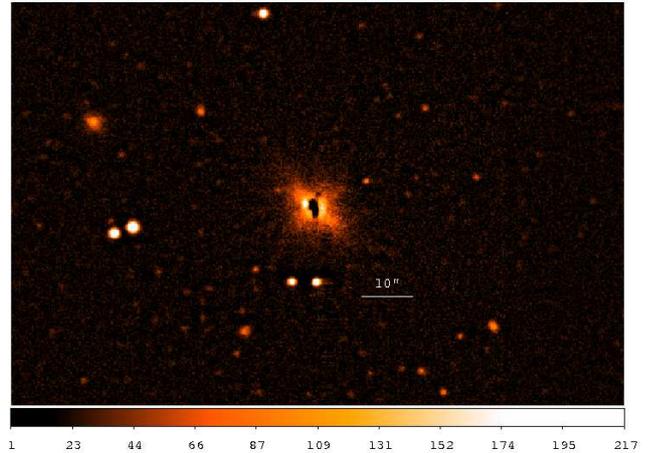}
\caption{Unsharp masked V band image obtained from optical data
  acquired at the ESO/MPI 2.2m telescope with the WFI; North is up,
  and East to the left. The white bar indicates $10''$ length. The
  inner dust lane at the center of NGC~3311 is about $2''$ wide and
  $7''$ long \citep{laine+03}, and is embedded within a central light
  excess of about $16''$ in diameter, where bright knots are visible
  here and in the HST image of \cite{laine+03}.}
\label{V+us}
\end{figure}

As a first step in the study of the light distribution in NGC~3311, we
determine the extension of the dust lane in its central region. We use
the {\it FMEDIAN} task in IRAF with a smoothing box of $15 \times 15$
pixels, and compute the ratio of the V band image and its {\it
  FMEDIAN} smoothed version; the resulting unsharp masked V-band image
is shown in Figure~\ref{V+us}.

This figure illustrates the presence of several distinct components at
the center of NGC~3311. A complex dust lane crosses the galaxy center
in the direction North-South (NS)
in the high angular resolution image in \cite{laine+03}.
Bright knots are seen East and Southwest of the galaxy center, within
and around the dust lane out to $8''$ in radius. According to
\cite{vast91}, the dominant knot is bluer, $\Delta (B-r) = -0.10$,
than the surrounding stellar population.

\section{Surface photometry for NGC~3311 in V 
  band}\label{2dphot}

In this Section, we describe isophotal analysis for the light
distribution of NGC~3311 from the V-band data. We then derive V-band
surface brightness profiles along several specific directions through
the galaxy to characterize the deviations from a simple concentric
elliptical light distribution.

\subsection{Isophote fitting}\label{isofit}
\begin{figure}[hbt!] 
\includegraphics[width=9.5cm]{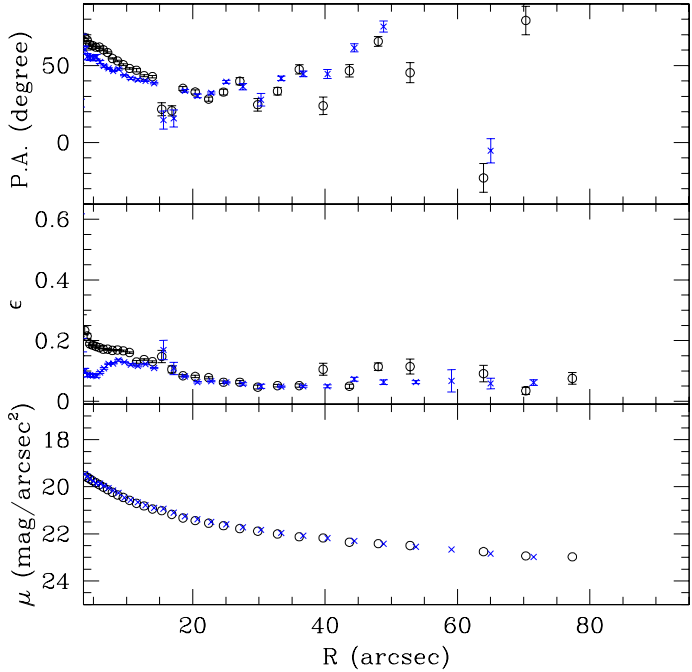}
\caption{One dimensional profiles of the P.A. for the isophote major
  axis, ellipticity, and surface brightness obtained with {\it
    ELLIPSE} from V-band images of NGC~3311 (VLT/FORS1 blue crosses,
  2.2m/WFI black open dots). {Because of the central dust lane,
    the profiles are well defined only outside $3''.5$.} }
\label{ellipseVK}
\end{figure}

We use the {\it ELLIPSE} task in IRAF on the V images to
perform the isophotal analysis of NGC~3311. Position angle (P.A.),
ellipticity, and average surface brightness profiles for NGC~3311 are
shown in Figure~\ref{ellipseVK}.

The {\it ELLIPSE} V-band average surface brightness profile extends
out to $80''$ from the galaxy center before the halo light from
NGC~3309 becomes significant. Note the excellent agreement between the
VLT/FORS1 and WFI surface brightness profiles.  The total integrated
magnitudes inside circular apertures, whose radius corresponds to the
last measured point, are: $m_{V}^{WFI}(R=77''.4)=11.55$ mag;
$m_{V}^{FORS}(R=71''.5)=11.36$ mag.  The half-light radii evaluated
from the light-growth curves out to the last measured point are
$R_{g,V}^{WFI} = 36'' \pm 2''$ ($\sim 8.9$ kpc), $R_{g,V}^{FORS} =
32'' \pm 2''$ ($\sim 7.9$ kpc).

At $R \le 5''$, the presence of a dust-lane is signaled by the large
variations in the V-band ellipticity and P.A. profiles.  Between $5''$
and $15''$ the V-band isophotes twist by about 20 degrees: this is
caused by the presence of the inner bright knots seen around the dust
lane (see Sect.~\ref{Vband}).  A twist of the NGC~3311 isophotes is
also reported by \cite{vast91}, but the P.A. variation was not
quantified.  For radii $15''<R < 45''$, the V-band ellipticity and
P.A. profiles are nearly constant, with average values $\epsilon
\simeq 0.05$ and P.A.$ \simeq 32^{\circ}$, i.e., the
isophotes in this region are nearly round.

\subsection{Analysis of surface brightness profiles along 
principal  axes of the light distribution}\label{profiles}

\begin{figure}[hbt!] 
\includegraphics[width=8.5cm]{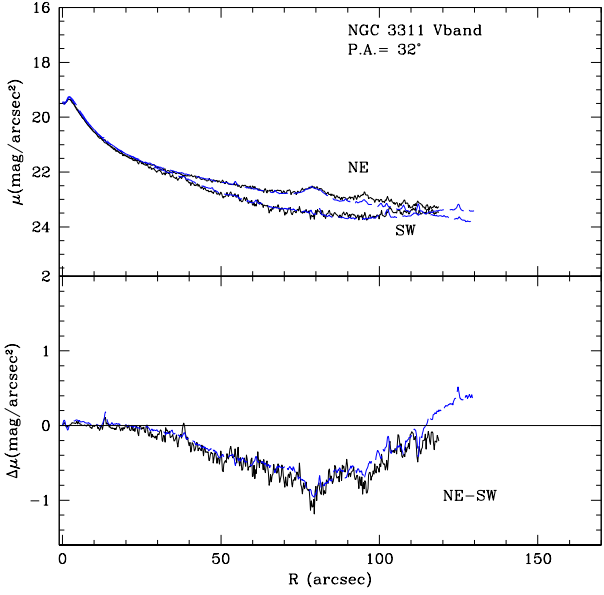}
\caption{Upper panel: folded V-band profiles extracted along
  P.A.$=32^\circ$ (major axis, 2.2m/WFI black full lines, VLT/FORS1
  blue dashed lines). NE is along P.A.$=32^\circ$ and SW along
  P.A.$=212^\circ$.   
  Lower panel: difference profiles.  The folded profile along the
  major axis illustrates the excess of light in the NE
  quadrant of the NGC~3311 halo, in the range of radii $20'' < R <
  120''$, with a maximum excess of about one magnitude at $R = 80''$.}
\label{foldp32}
\end{figure}

\begin{figure}[hbt!] 
\includegraphics[width=8.5cm]{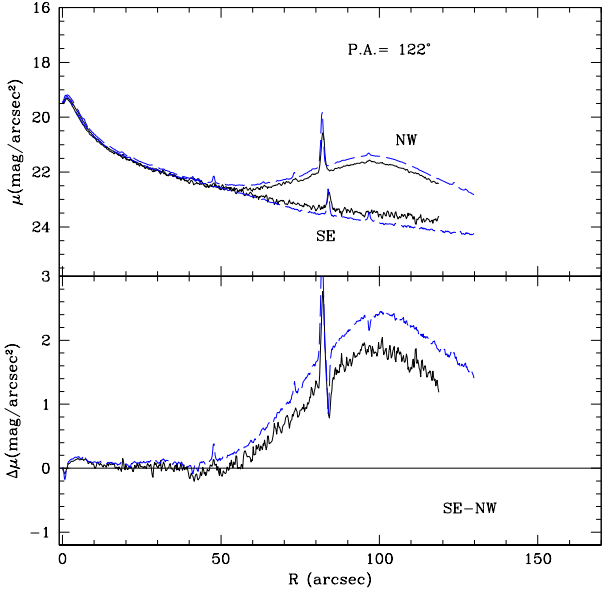}
\caption{Same as Fig.\ref{foldp32}. Upper panel: folded V-band
  profiles extracted along P.A.$=122^\circ$ (minor axis, 2.2m/WFI
  black full lines, VLT/FORS1 blue dashed lines).  SE is along
  P.A.$=122^\circ$, NW along P.A.$=302^\circ$.  Lower panel:
  difference profiles. On the NW side, for $R \ge 65''$ we see the
  contribution from the outer regions of NGC~3309.}
\label{foldp122}
\end{figure}

\begin{figure}[hbt!] 
\includegraphics[width=8.5cm]{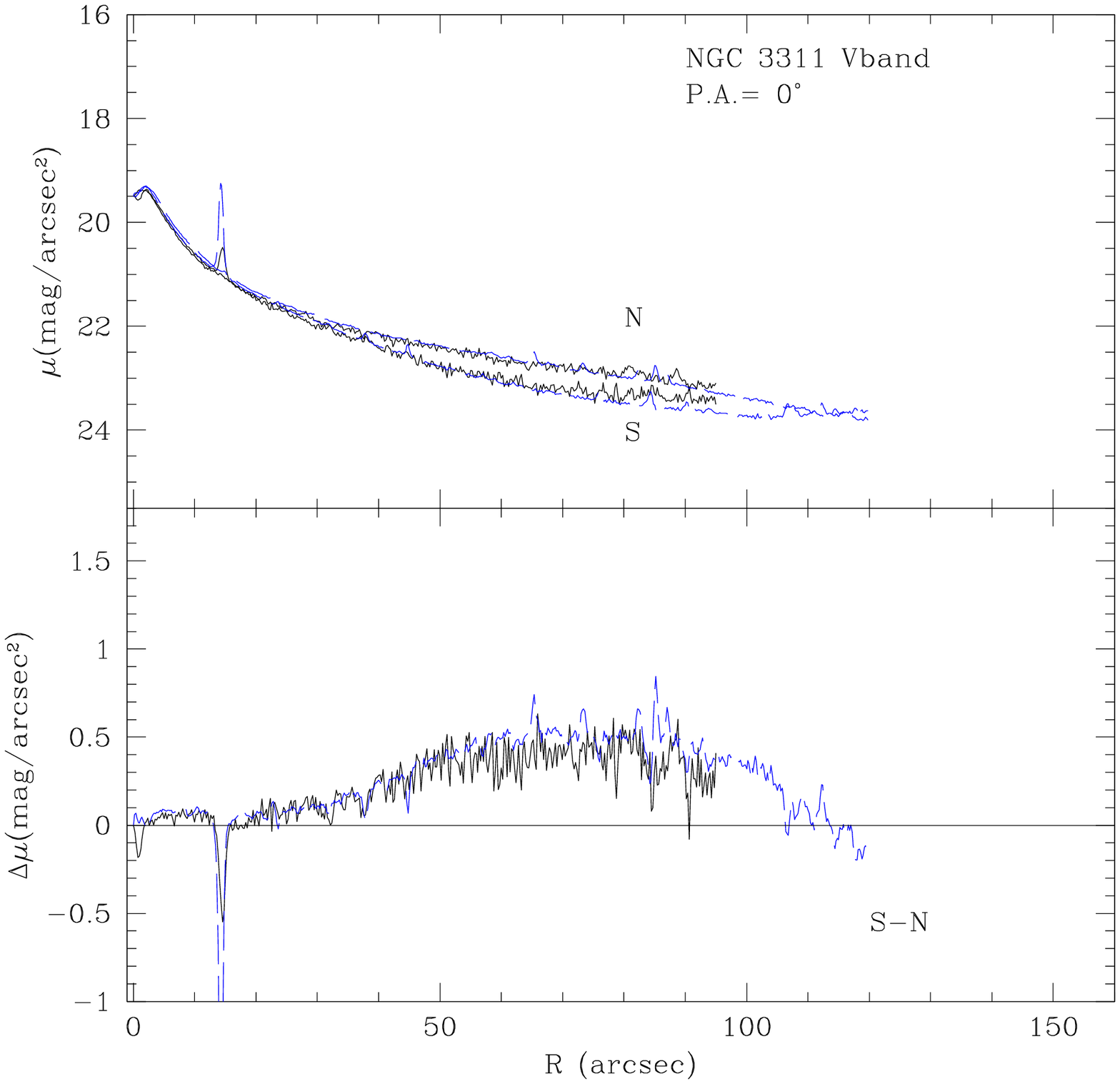}
\caption{Same as Fig.\ref{foldp32}. Upper panel: folded V-band
  profiles extracted along P.A.$=0^\circ$ (2.2m/WFI black full lines,
  VLT/FORS1 blue dashed lines).  N is along P.A.$=0^\circ$, S along
  P.A.$=180^\circ$.  Lower panel: difference profiles.}
\label{foldp0}
\end{figure}

\begin{figure}[hbt!] 
\includegraphics[width=8.5cm]{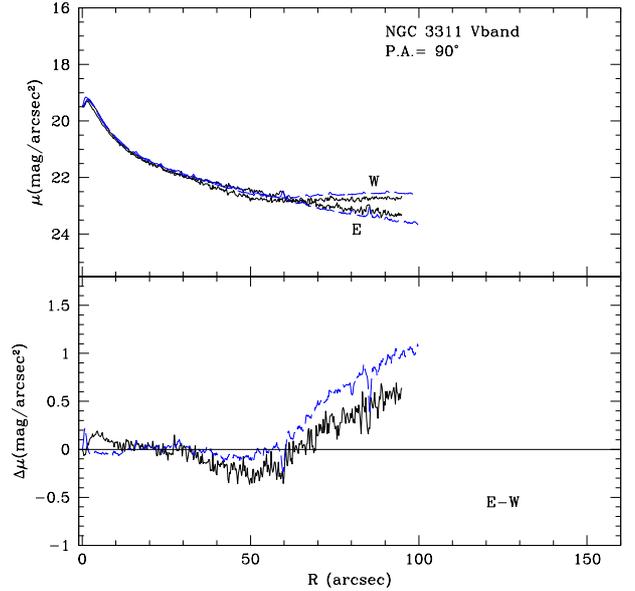}
\caption{Same as Fig.\ref{foldp32}. Upper panel: folded V-band
  profiles extracted along P.A.$=90^\circ$ (2.2m/WFI black full lines,
  VLT/FORS1 blue dashed lines).  E is along P.A.$=90^\circ$, W along
  P.A.$=270^\circ$.  Lower panel: difference profiles.}
\label{foldp90}
\end{figure}

The position angle profile from the isophote fit in
Fig.~\ref{ellipseVK} has a complicated radial dependence which we show
now is related to asymmetries in the light distribution. To this end
we investigate separately the surface brightness profiles along
several axes across the light distribution in NGC~3311. From the
isophote fitting, the major axis is at {P.A.$=32^\circ$ } and the
minor axis is at P.A.$=122^\circ$.  Along these directions, we have
extracted one-dimensional (1D) surface brightness profiles from V-band
VLT/FORS1 and 2.2m/WFI images.

The V band surface brightness profiles become asymmetric for $R>30''$,
as shown in Figs.~\ref{foldp32} and \ref{foldp122}.  The major-axis
profile at P.A.$=32^\circ$ (North East - NE) is up to one magnitude
brighter than that at P.A.$=212^\circ$ (South-West - SW) in the radial
range $ 20'' < R < 120''$ (Fig.~\ref{foldp32}).  There is additional
light along P.A.$=212^\circ$ at $R \ge 100''$ which may be related
with a substructure in the Hydra~I core. This will be further
investigated in the residual image, once the light distribution from
the bright galaxies is accounted for.  The light profiles along the
minor axis, at P.A.$=122^\circ$ (South East - SE) and P.A.$=302^\circ$
(North West - NW) are symmetric for $R < 60''$
(Fig.~\ref{foldp122}). At P.A.$=302^\circ$ and $R> 60''$ we see the
light from the NGC~3309 halo.  For $R \le 5''$, both V band profiles
show the absorption by the central dust-lane.

In addition, we extract folded profiles along the NS (P.A.$ =0^\circ$)
and EW (P.A.$=90^\circ$) directions. These are also asymmetric; see
Fig.~\ref{foldp0} and Fig.~\ref{foldp90}. At P.A.$=0^\circ$ and $R \ge
20''$, the galaxy becomes brighter than at P.A.$=180^\circ$; the
difference is about 0.5 mag at $R = 80''$ (Fig.~\ref{foldp0}). At
P.A.$=90^\circ, 270^\circ$ the profiles are symmetric for $R \le
30''$, while in the radial range $30'' \le R \le 60''$ the profile at
P.A.$=90^\circ$ is brighter (by about 0.2 mag at $R \sim 50''$) than
that at P.A.$=270^\circ$. At larger radii the profile is affected by
the light of NGC~3309. {Figs.~\ref{foldp0} and ~\ref{foldp90}
  illustrate that the additional light is distributed over an opening
  angle larger than $90^\circ$ as seen from the center of NGC~3311,
  therefore the substructure leading to the asymmetry in
  Fig.~\ref{foldp32} is not a radial stream or fan.}

The VLT/FORS1 and 2.2m/WFI profiles in Figs.~\ref{foldp32} and
\ref{foldp0} show very good agreement on both the N/NE and S/SW
sides of NGC~3311 and also in the difference profiles.  This shows
that systematic effects caused by scattered light from the bright star
NE of NGC~3311 are negligible and that there are no residual gradients
due to the sky background along these directions.
Figs.~\ref{foldp122} and \ref{foldp90} show a small relative variation
between the W/NW and E/SE sides, amounting to $\sim 0.2$ mag over
$200''$.  As we will discuss further in Sect.~\ref{1dplus2d},
this does not affect our main results.

The analysis presented in this Section illustrates that the halo of
NGC~3311 is asymmetric. Between P.A.  $=0^\circ$ and P.A.$=90^\circ$,
and in the radial range $30''<R < 120''$, it is brighter than in the
remaining three quadrants.

\section{Models for the light distribution in NGC~3311, 
  NGC~3309, and  the residual diffuse light in the Hydra I 
 cluster core}\label{symmodels}

In this Section we construct two dimensional (2D) surface brightness
models for the light distribution in the Hydra cluster core around
NGC~3311. Any such model must include the light distribution of
NGC~3309, because the outer regions of the two elliptical galaxies
overlap along the LOS.  Furthermore, the modeling must also account for the
asymmetric bright component in the NE outer region of NGC~3311,
signaled by the asymmetry of the extracted profiles in
Section~\ref{profiles}.

\subsection{V-band 2D model for NGC~3311 and
  NGC~3309}\label{2dgalfitv}

We model the surface brightness distribution of NGC~3311 and
NGC~3309 from the V band VLT/FORS1 image. For fitting the model to the
VLT/FORS1 data, we first adopt the mask shown in the middle panel of
Fig.~\ref{VFORS1}, identified as ``cmask'' in what follows.  It masks
the saturated/bright stars in the field, the central $R< 5''$ of
NGC~3311, a region $R < 8''$ around the saturated center of
NGC~3309, and the background galaxies.

For the ``cmask'' model fit, the 2D Sersic model has parameters $n_{V}
= 3.3 $, $R_e=21''.1$ for NGC~3309 and $n_{V} = 5.0$, $R_e=270''.6$
for NGC~3311; see Tables~\ref{galfit3309} and \ref{galfit3311}.  The
residual V band image {shows a number of structures}. In
particular, to the SW of NGC~3311 around the galaxy's major axis at
P.A.$= 212^\circ$, in a $\sim 90^\circ$-wide cone and radial range $
40'' < R < 80'' $, we see (unphysical) negative residuals of about
$260$ ADU per pixel with respect to the sky counts measured in the
``empty'' eastern boundary of the VLT/FORS1 image. This negative
residual signals that NGC~3311 is fainter than the 2D Sersic model
there. {It derives from fitting an azimuthally averaged model to
  an intrinsically asymmetric light distribution; we saw previously in
  Sect.~\ref{profiles} that the galaxy is brighter in the NE quadrant
  than in the other three quadrants.}

\begin{figure*}[hbt!] \centering
\includegraphics[width=16.5cm]{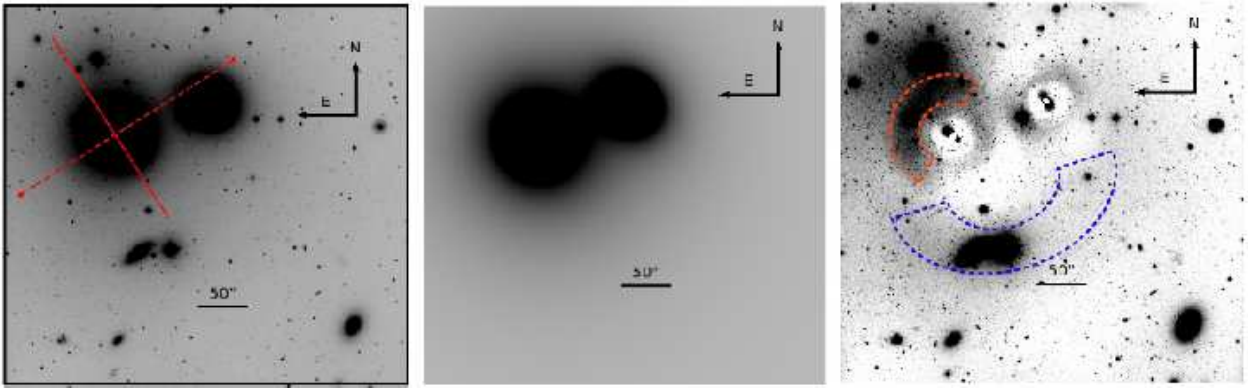}  
\caption{GALFIT model of the VLT/FORS1 image obtained with ``allmask''
  (see right panel of Fig.~\ref{VFORS1}). Left panel: VLT/FORS1 V band
  image. Full and dashed red lines indicate the major axis at
  PA=$32^\circ$, and the minor axis, respectively.  Center panel:
  combined 2D GALFIT model for NGC~3311 and NGC~3309. Right panel:
  residual image, {showing extra light associated with the NE
    outer halo of NGC~3311, a region of negative residuals $\sim 60''$
    South West of NGC~3311, and preliminary evidence for two tidal
    streams around HCC~026 and HCC~007, outlined by the red- and blue
    dashed curves.} The image size in all panels is
  $6'.8\times\,6'.4$, and darker colors indicate brighter regions.  }
\label{v2dmod}
\end{figure*}

Therefore, we next investigate whether by masking the whole North-East
quadrant of NGC~3311, the ``symmetric'' 2D model for the V band light
in NGC~3311 will acquire a lower value for the Sersic index $n$, and
thus the problem of negative residuals at P.A.$= 212^\circ$ is
reduced. We define a new mask which additionally covers most of the NE
quadrant in the NGC~3311 halo: this is shown in the right panel of
Fig.~\ref{VFORS1} and is identified as ``allmask''.  When the light
excess is masked, the GALFIT 2D fit to the V band FORS1 image returns
a Sersic index $n_{V} = 4.8$ and $R_e = 198''.8$ for NGC 3311.  The
GALFIT model of NGC~3309 and NGC~3311, using ``allmask'' with the
VLT/FORS1 data, and the residual image (difference between image and
model) are shown in Figure~\ref{v2dmod}. Parameters are given in
Tables~\ref{galfit3309} and \ref{galfit3311}.

The residual image still shows negative values in the same region
in a $\sim 90^\circ$-wide cone around P.A.$= 212^\circ$ and for the
radial range $ 40'' < R < 80'' $, but the level is now reduced to 110
ADU per pixel, compared to 260 ADU per pixel in the ``cmask'' residual
image. Therefore the extra light in the NE of NGC 3311 must extend
slightly further than the masked area.  

In addition, the residual image shows several positive structures
which correspond to additional light in the real image with respect to
the 2D models of NGC~3311 and NGC~3309, at a level of 200-500 ADU per
pixel relative to sky. {These structures will be more quantitatively
  discussed in Sects.~\ref{1dplus2d} and \ref{Ltails} in the context
  of a ``maximal'' symmetric model with which the negative residuals
  are removed. Briefly,}

\begin{itemize}
\item In the outer regions, NGC~3311 is more luminous than the best
  fit 2D model in a wide area covering the whole NE quadrant to the
  galaxy.  Superposed on this envelope we also see a prominent tidal
  stream around HCC~026, one of the dwarf galaxies in the Hydra~I
  cluster core. The stream is outlined by the red dashed lines in
  Fig.~\ref{v2dmod} around $\sim 43''$ distance from the NGC~3311
  center. The tidal tails around HCC~026 extend about $15$ kpc NW and
  $5$ kpc NE of the dwarf galaxy.

\item The S0 galaxy HCC~007 situated $119''$ South of NGC~3311 appears
  to be embedded in an extended, thick, tidal stream visible at faint
  surface brightness levels.  Emerging from HCC~007, the stream
  circles the south part of the Hydra~I core, it encompasses both
  NGC~3311 and NGC~3309, and extends over at least $450''$ ($\sim 110$
  kpc), as indicated by the dashed blue lines in
  Fig.~\ref{v2dmod}. The eastern tail of HCC~007 appears to join the
  bright envelope to the NE of NGC~3311; as part of the envelope it
  may continue considerably further to the north.
\end{itemize}

\begin{table}
\begin{tabular}{c c c c}     
\hline\hline       
Parameter & V band & V band & V band \\ 
&  FORS1 & FORS1 & FORS1 \\
&        &       & max.~symmetr.  \\
\hline                    
Comp. type  & Sersic & Sersic & Sersic \\
$m_{tot}$     & $11.6\pm 0.05$  & $11.6 \pm 0.05$ & $11.2 \pm 0.01$ \\  
R$_e$      & $21''.1 \pm 0''.07$ & $21''.9 \pm 0.07$ & $35''\pm 0.25''$ \\
n           & $3.3 \pm 0.02$ & $3.3 \pm 0.02$ & $5.7 \pm 0.04$ \\
b/a        & $0.87 \pm 0.01$  & $0.85 \pm 0.01$ & $0.87 \pm 0.01 $ \\
P.A.        & $48.2 \pm 0.2$ & $50.7 \pm 0.2$ & $49.8 \pm 0.2$ \\
mask &  ``cmask''&  ``allmask'' &  ``allmask''\\

\hline                  
\end{tabular}
\caption{Parameters for the 2D GALFIT models of
  NGC~3309. Apparent V band total magnitudes are in 
agreement with the published NED values (RC3).}\label{galfit3309}
\end{table}

\begin{table}
\begin{tabular}{c c c c}     
\hline\hline       
Parameter  & V band & V band & V band \\  
  &  FORS1  & FORS1 & FORS1 \\
 &    &       & max.~symmetr. \\
\hline 
Comp. type  & Sersic & Sersic & Sersic \\  
 $m_{tot}^*$  & $9.7\pm0.01$ & $10.1\pm0.01$ & 9.52 \\ 
R$_e^*$  & $270''.6 \pm 3''.4$ & $198''.8 \pm 2''.2$ & $850'' \pm 50''$ \\ 
n  & $5.0 \pm 0.02$ & $ 4.8 \pm 0.02 $ & $ 10.5 \pm 0.1$\\ 
b/a  & $0.93 $  & $0.93 $ &$0.93$ \\ 
P.A.  & $38\pm 0.5$ & 32 & 32 \\ 
mask  &  ``cmask'' & ``allmask'' & ``allmask''\\
\hline
\end{tabular}
\caption[]{Parameters for the 2D GALFIT fit models of
  NGC~3311.\\ {$^*$ The $m_{tot}$ and R$_e$ values computed from the 
    GALFIT fit model for NGC 3311 are
    respectively brighter and larger than those computed in Sect.~\ref{isofit} 
    from the V band light growth curve, because the latter is measured only out 
    to $71''.5$, while the 2D model is fit to the whole
    $6'.8 \times 6'.4$ area.}}\label{galfit3311}
\end{table}

\begin{figure}[hbt!] 
\includegraphics[width=9.5cm]{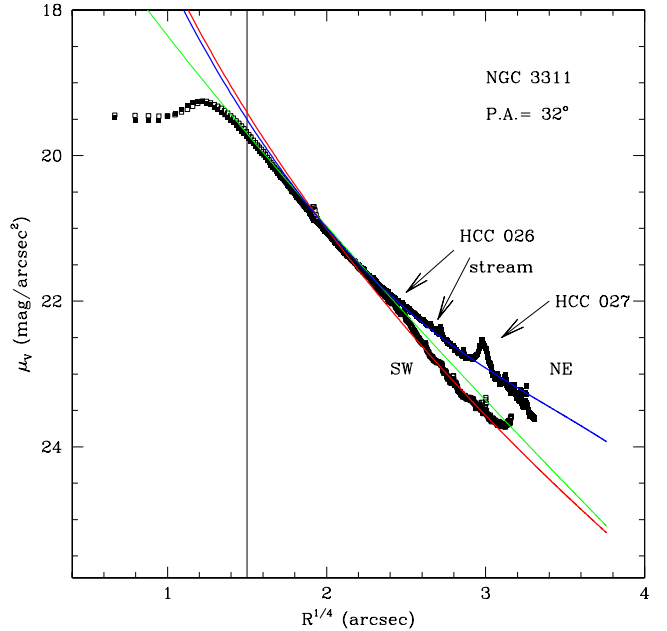}
\caption{Folded 1D surface brightness profile along the major axis of
  NGC~3311 plotted as function of $R^{1/4}$. Open symbols show values
  along P.A.$= 212^\circ$ (SW), full symbols those along P.A.$=
  32^\circ$ (NE of the galaxy center). The green continuous line shows
  the extracted 1D major axis profile of the Sersic ``allmask'' model
  for the V-band FORS1 data described in Sect.~\ref{2dgalfitv}. The
  red continuum line is the major axis profile for the light centered
  on NGC~3311 (``maximal'' symmetric model, MSM).  The blue full line
  shows the 1D fit to the surface brightness profile along P.A.$=
  32^\circ$ (NE) with a two component model, consisting of the MSM
  plus a one-sided exponential profile . The arrows indicate the
  position of the stream around HCC~026 and the light at $R\simeq
  80''$ associated with the dwarf galaxy HCC~027 which falls on the
  extracted profile. See Sect.~\ref{1dplus2d} for details. }
\label{1DVprofs}
\end{figure}

Fig.~\ref{1DVprofs} shows 1D surface brightness profiles extracted
along the NE and SW sides of the major axis of NGC 3311, as
illustrated in Fig.~\ref{v2dmod}. For radii $ 5'' < R < 30'' $, the
extracted profiles are straight lines in the $R^{1/4}$ plot: deviations
from a de Vaucoleurs profile become large at $R > 30''$ on the NE side
(P.A.$= 32^\circ$), indicating additional light there. 
The flattening of the surface brightness profile at P.A.$=
32^\circ$ in Fig.~\ref{1DVprofs} with respect to the steeper profile
at P.A.$= 212^\circ$ indicates that at radii larger than $30''$, the
outer surface brightness contours shift towards the NE; i.e., the
envelope is off-centered to the NE relative to the bright central
region of NGC~3311.  This confirms the asymmetries of the extracted
surface brightness profiles discussed in Sect.~\ref{2dphot}.

Fig.~\ref{1DVprofs} also explains the origin of the negative residuals
in the ``allmask'' model: the surface brightness profile fitted by
GALFIT ({green} line) falls slightly above the SW (P.A.$=212^\circ$)
major axis profile because of the surface brightness drop around 30
arcsec, so that the symmetric 2D model built from this profile
generates negative values in the residual image about P.A.$=
212^\circ$.

In the next Section, we will improve on this by constructing a
``maximal'' symmetric model for the light centered on NGC~3311.
Then we will use this to construct a new 2D model, and to discuss
the structures in the corresponding residual image in more detail.

\subsection{The asymmetric light distribution around NGC~3311:
  extended off-centered envelope and tidal tails}\label{1dplus2d}

We build a maximal symmetric model (MSM) for NGC~3311 in two steps.
First, we fit the steep surface brightness profile towards SW, along
P.A.$= 212^\circ$, with a Sersic law.  {Second, we build a 2D
  GALFIT model with the Sersic parameters fixed by the 1D fit to the
  steep profile, which therefore leads to positive or null residuals
  everywhere in the NGC~3311 halo.}  

Fig.~\ref{1DVprofs} shows the 1D fit profile over-plotted on the data
(red line); the model fits the data well in the radial range
$15''$-$80''$. The parameters are $n_V=10.5 \pm 0.1$, $R_e = 850''\pm
50''$, and $\mu_e = 28.07 \pm 0.1 $ mag arcsec$^{-2}$; see
Table~\ref{galfit3311}.  Now we use GALFIT with the mask ``allmask''
to generate a 2D MSM for NGC~3311, using the Sersic law with
parameters set from the 1D best fit, and the major axis PA=$32^\circ$
and axis ratio $0.93$ from our previous 2D model fits. Then, keeping
this component fixed, we use GALFIT again to determine new best fit
parameters for NGC~3309, matching a Sersic model to the remaining
light. Results of this NGC 3309 fit are listed in
Table~\ref{galfit3309}. Adding both components, we obtain the MSM for
the two bright galaxies NGC~3311 and NGC~3309.  By construction, the
residual image obtained as the difference between the V-band VLT/FORS1
image and this MSM should show only positive or null residual values
around NGC~3311 (outside the central $R < 15''$ containing dust and
bright knots).

\begin{figure*}[hbt!] \centering
\includegraphics[width=14.0cm]{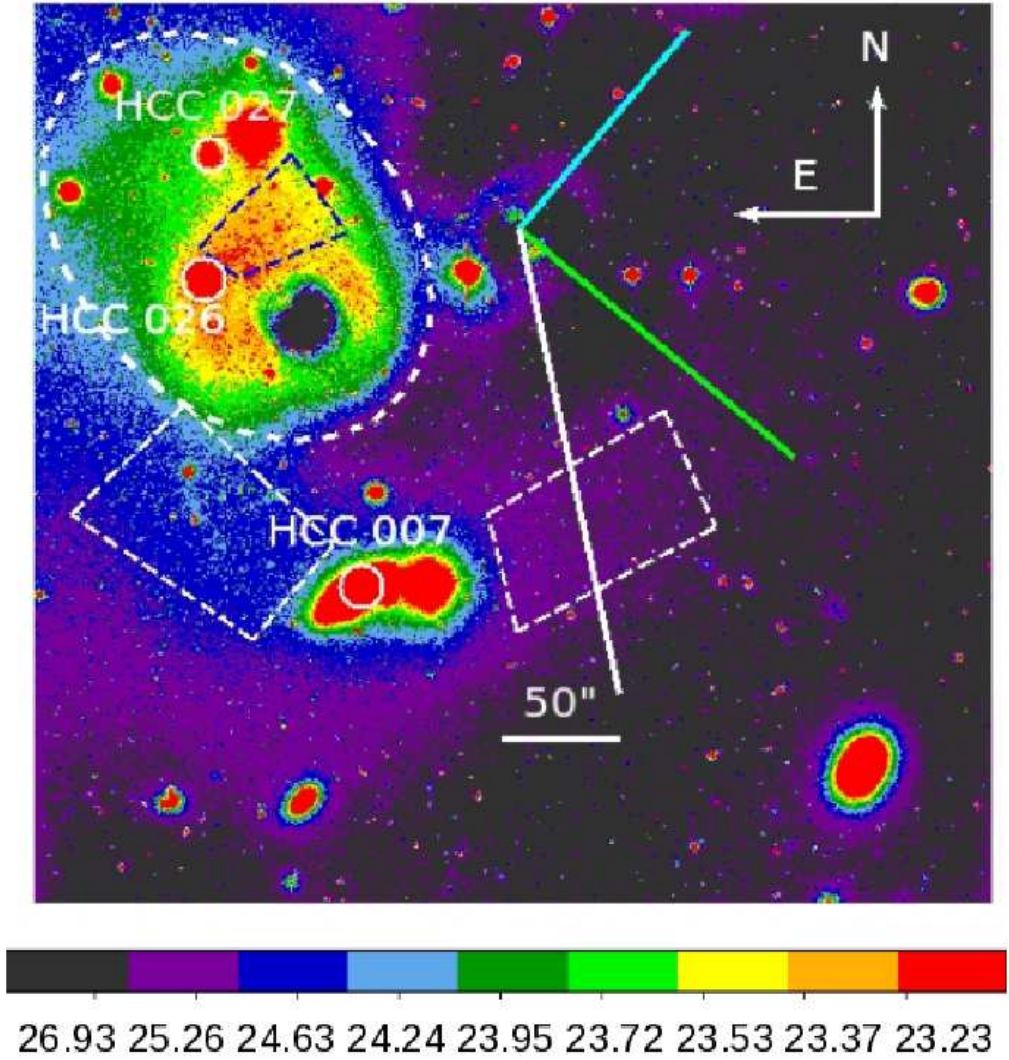}
\caption{Residual VLT/FORS1 V-band image with respect to the MSM for
  NGC~3311 and NGC~3309. The white ellipse indicates the extended halo
  off-centered towards the NE of the bright inner regions of NGC~3311
  {(black). The angular extent of this envelope as seen from the
    center of NGC~3311 is consistent with the opening angle of the
    additional light as shown by Figs.~\ref{foldp0} and ~\ref{foldp90}
  }.  The most noticeable substructure on top of the extended envelope
  NE of NGC~3311 is the tidal stream associated with the dwarf galaxy
  HCC~026. Its NW part is indicated on the image with the blue dashed
  polygon.  At larger radii, and at fainter surface brightness levels,
  we see a thicker tail around HCC~007; the luminosity in this tail is
  computed in the areas limited by the white polygons, see
  Sect.~\ref{Ltails}. {The two small objects near the center of
    the NE polygon appear to be connected by a further small stream};
  see also Fig.~\ref{v2dmod}.  The FoV is $6'.8 \times 6'.4$. {
    Colour levels are labeled with the corresponding surface
    brightness values.  This image shows that in the region south of
    HCC~007 the background is flat, with a systematic E-W gradient
    across the image at less than $\sim 10$ ADU.  The white, green,
    and blue lines from the center of NGC~3309 indicate the P.A.s at
    $= 190^\circ, 230^\circ\, \mbox{and}\, 320^\circ$ respectively;
    see extracted profiles in Fig.~\ref{N3309prof}}.}
\label{VresFORS1WFI}
\end{figure*}

This residual image, displayed in Fig.~\ref{VresFORS1WFI}, clearly
shows the presence of an additional, extended envelope, off-centered
towards the NE of the inner regions of NGC~3311. It is highlighted by
the white dashed ellipse in Fig.~\ref{VresFORS1WFI}. The residual
image also shows the bright tail around HCC~026, and the broader tail
around HCC~007 which connects with the off-centered envelope. {
  These structures stand out more clearly in Fig.~\ref{VresFORS1WFI}
  than in the earlier Fig.~\ref{v2dmod} because with the MSM we no
  longer have negative residuals. Note that the ``half-ring''
  morphology around NGC 3311 slightly inside the HCC~026 stream
  probably signals that a small fraction of the light in the
  ``maximal'' symmetric inner halo for $r<30''$ should be counted as
  part of the off-centered envelope, such that its surface brightness
  on top of NGC 3311 becomes a constant $\mu_V=23.4$ mag
  arcsec$^{-2}$. However, we have no independent way to determine this
  light fraction. The ``maximal'' symmetric model is constructed such
  as to put the minimal fraction of light into the off-centered
  envelope.  }

To quantify the asymmetry in the NGC 3311 halo, we also fit the NE
surface brightness profile in Fig.~\ref{1DVprofs} (P.A.$= 32^\circ$)
with a two-component model, consisting of a Sersic law with the same
$n$ and $R_e$ and an additional exponential profile for the outer
regions. The exponential profile often is a good approximation for the
outer stellar envelopes of cD galaxies \citep{seigar+07}. This
combined model provides a good fit to the NE profile which is also
shown in Fig.~\ref{1DVprofs} (blue line); the parameters are $n_V=10.5
\pm 0.1$, $R_e = 850''\pm 50''$ and $\mu_e = 28.2 \pm 0.1 $ mag
arcsec$^{-2}$ for the Sersic component, and $\mu_0=23.2\pm 0.1$ mag
arcsec$^{-2}$ and $r_h = 200''\pm 50''$ for the exponential component.
The peak surface brightness of the off-centered envelope { in the
  ``maximal'' symmetric halo model} is $\mu_{0,V} = 23.2$ mag
arcsec$^{-2}$, as given by the exponential fit along P.A.$= 32^\circ$.

{In the residual map, Fig.~\ref{VresFORS1WFI}, the measured
  surface brightness at the position of the HCC~026 tail is also $\sim
  23.2$ mag arcsec$^{-2}$. This is the combined surface brightness of
  the off-centered envelope and the tail, which are superposed along
  this line-of-sight. The contribution of the HCC026 tail within the
  polygon area limited by the blue dashed line in
  Fig.~\ref{VresFORS1WFI} is about 23\%; see the next section.  Hence
  the average surface brightness of the tail around HCC~026 in the
  area enclosed by the blue dashed polygon in Fig.~\ref{VresFORS1WFI}}
is $\mu_V = 24.8 \pm 0.2$ mag arcsec$^{-2}$.  

The surface brightness of the tail around HCC~007 can be directly
measured on the residual image. It is brighter on the NE side where
$\mu_V = 24.4 \pm 0.5$ mag arcsec$^{-2}$, than on the NW side where it
is about one magnitude fainter. {The brighter NE tail of HCC~007
  is clearly a localized structure with higher surface brightness than
  other areas at similar distance from either NGC 3311 or NGC
  3309. The central region of the brighter NE tail of HCC~007 contains
  two small objects that appear to be connected by a possible further
  small stream.  }

\begin{figure}[hbt!] 
\includegraphics[width=9.5cm]{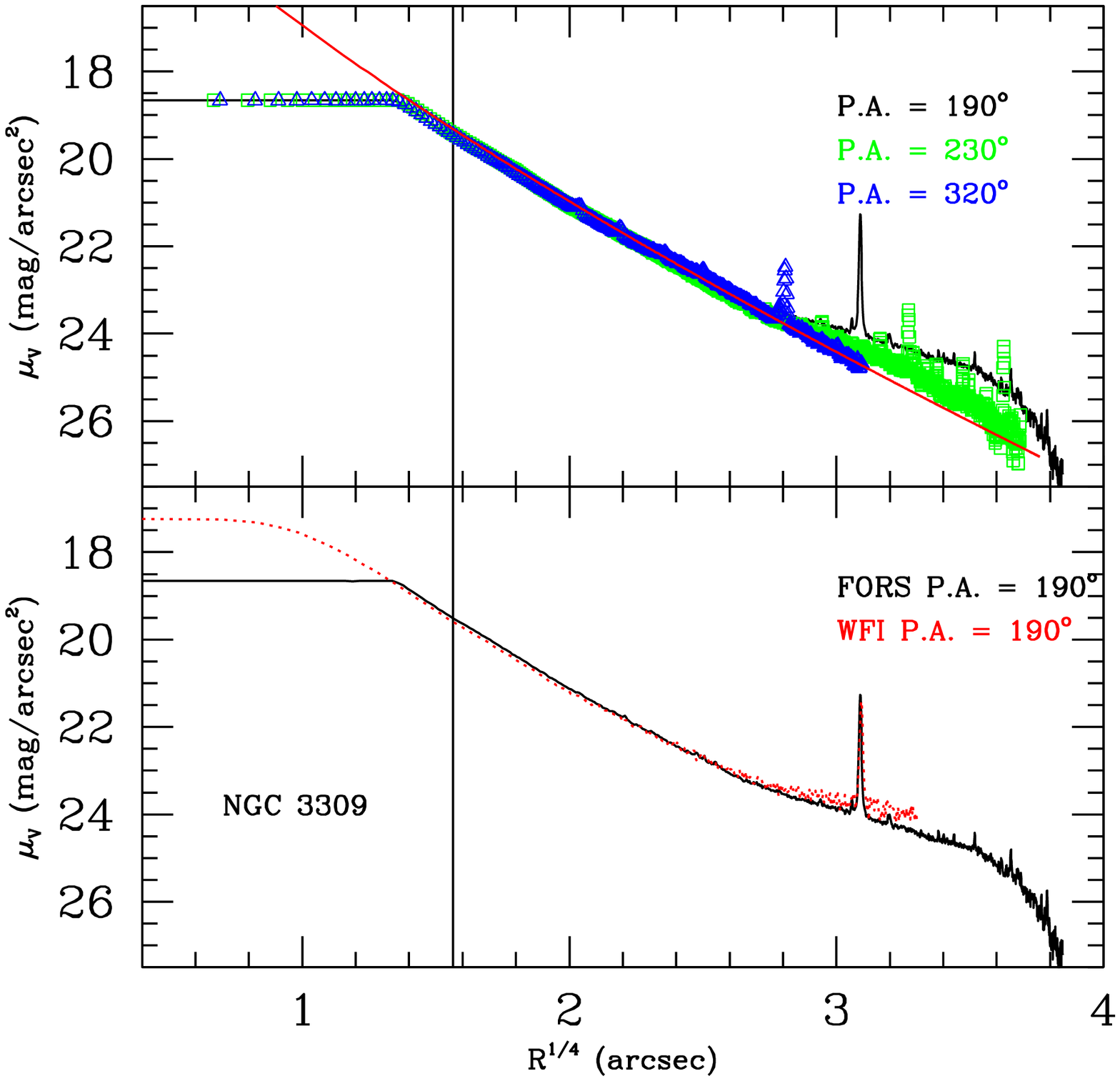}
\caption{Upper panel: surface brightness profiles extracted in
  $5^\circ$ wide cones along the major (P.A.$= 230^\circ$, green
  squares), minor axis (P.A.$= 320^\circ$, blue triangles) and
  intermediate axis (P.A.$=190^\circ$, black line) of NGC~3309,
  plotted as function of $R^{1/4}$. {The P.A. at $ 190^\circ,
    230^\circ\, \mbox{and}\, 320^\circ$ are indicated by the white,
    green, and blue lines from the center of NGC~3309 in
    Fig.~\ref{VresFORS1WFI}}. The red continuous line shows the
  extracted 1D major axis profile for the 2D GALFIT ``maximal''
  symmetric model (MSM) to the light of NGC~3309. While this is a good
  fit for the minor axis profile at large radii, profiles along P.A.$=
  230^\circ$ and P.A.$=190^\circ$ show additional light, as they go
  trough the brightest part of the HCC~007 NW stream. Lower
  panel: surface brightness profile from VLT/FORS1 (black line) and
  2.2/WFI (red line) V band images along P.A.$=190^\circ$. }
\label{N3309prof}
\end{figure}

We can check with Figure~\ref{N3309prof} that the NW tail of HCC~007
is not an artifact of residual halo light in NGC~3309. The upper panel
of this figure shows that the minor axis profile of this galaxy
(P.A.$= 320^\circ$) is well described by the fitted Sersic profile,
whereas additional light is seen along P.A.$= 230^\circ$ and
particularly along P.A.$= 190^\circ$ which goes through the brightest
part of the NW tail close to HCC~007. {However, it cannot be ruled
  out that this region contains some light from a very diffuse, low
  surface brightness structure around NGC 3311 and the cluster
  center.}

Finally, we show that the morphologies of the substructures seen in
the residual image in Fig.~\ref{VresFORS1WFI} are not due to
systematic effects in the 2D modeling caused by uncertainties in the
background estimates.  This concern can be addressed by showing that
these substructures are present in two independent datasets, the
VLT/FORS1 and 2.2m/WFI images.  We have already seen in
Sect.~\ref{profiles} that the asymmetries in the extracted profiles
along P.A.$=32^\circ,212^\circ$ are very similar in the two data sets;
this confirms independently the presence of the off-centered envelope
NE of NGC~3311.  The lower panel of Fig.~\ref{N3309prof} compares the
VLT/FORS1 and WFI surface brightness profiles through the fainter NW
tail of HCC~007 along P.A.$= 190^\circ$ as seen from the center of
NGC~3309.  The light excess in the HCC~007 tail is clear in both data
sets.  Both results are independent of the slight residual gradient
described in Sect.~\ref{profiles}, because this is essentially
perpendicular to P.A.$=32^\circ$ and P.A.$=190^\circ$.  Finally, we
have also carried out 2D GALFIT modeling on the 2.2m/WFI V band image,
using the 1D maximal model fit parameters for NGC~3311, and the
``allmask'' mask shown in Fig.~\ref{Vfinal} to determine the
parameters of NGC~3309.  The residual V-band 2.2m/WFI image has a
lower S/N than that based on the VLT/FORS1 data.  Nonetheless both
elongated streams around HCC~026 and HCC~007 are independently
confirmed. In the case of HCC~007, only the brighter NE tail is
visible in the residual image.

\subsection{Luminosities of the substructures in the Hydra~I
  core}\label{Ltails}

We next compute the total luminosity of the off-centered envelope from
the flux within the elliptical aperture shown in
Fig.~\ref{VresFORS1WFI}. { We mask stars, dwarf galaxies, and a
  circular region with $R=25''$ centred on NGC~3311}. The resulting
luminosity in the off-centred envelope is $L_{\rm V,NE,env}= 1.2
\times 10^{10}\, (\pm 6.0 \times 10^8)\,L_\odot$\footnote{The
  $1-\sigma$ error in the luminosity is estimated by integrating the
  sky rms per pixel over the area enclosed by the elliptical
  aperture.}, using a distance of $50$ Mpc for the Hydra I
cluster. This estimate does not include the luminosity of the MSM halo
in this region, but it does include the light in the stream around
HCC~026, which is seen in addition to the exponential component in
Fig.~\ref{1DVprofs}.

For comparison, we compute the luminosity of the MSM {\it i)} within
the same elliptical aperture, and {\it ii)} in the circular annulus
between $R=25''$ and $120''$, by integrating the flux of the GALFIT
model output image over the respective areas.  The luminosities of the
MSM model evaluated in these two regions are $L_{\rm V,NE,halo} = 2.3
\times 10^{10}\, (\pm 6.0 \times 10^8)\,L_\odot$ and $L_{\rm V,halo} =
7.5 \times 10^{10}\, (\pm 2.0 \times 10^9)\,L_\odot$. Thus, within the
area limited by the elliptical aperture in Fig.~\ref{VresFORS1WFI},
the luminosity in the off-centered envelope amounts to 50\% of the
light in the MSM, while it is only 15\% of the total luminosity of the
MSM integrated between $25''$ and $120''$.

To estimate the luminosity of the HCC~026 tail we carry out surface
photometry in a polygon designed to include most of the NW half of the
stream; it is constructed with the {\it IRAF POLYMARK} and {\it
  POLYPHOT} tasks and is shown by the blue dashed polygon in
Fig.~\ref{VresFORS1WFI}.  { The polygon defined for the HCC~026
  tail is at an average distance of $ 53''$.} From the luminosity
measured in this polygon we subtract the contribution of the smooth
envelope in the same area, { estimated from the average residual
  surface brightness on both sides of the stream. This gives us a {\sl
    differential} measurement of the luminosity of the HCC~026 tail in
  the polygon.}  The resulting luminosity of the NW part of the stream
is $L_{\rm V,NW stream HCC~026}= 4.8 \times 10^{8} (\pm 8\times10^7)
L_\odot$, which is equivalent to an apparent magnitude of $m_{\rm
  V,stream HCC~026} = 16.6$.  {We find that the NW stream of HCC
  026 contributes about $ \sim 15\%$ of the combined surface
  brightness of the MSM halo and off-centered envelope at the stream
  position (see also Fig.~\ref{1DVprofs}), consistent with the
  estimate obtained independently from stellar population analysis in
  \citet{Coccato11b}}.  The luminosity in the southern part of the
HCC~026 stream is more difficult to disentangle from the diffuse halo,
but comparable.  {It is seen in Fig.~\ref{foldp90} as an excess of
  surface brightness of $\sim 0.1$ mag over $\sim 15''$ in the E
  profile over that on the W side.} The total stream luminosity is
clearly several times larger than the total luminosity of the dwarf
galaxy HCC~026 itself, $L_{\rm V,HCC~026} = 1.5 \times 10^8 L_\odot$
\citep{Misgeld08}, and if the stream is physically related to HCC~026,
a large part of the luminosity of this galaxy has already been tidally
dissolved!

In a similar way, we compute the light in the tail emerging from
HCC~007.  We define two polygons shown by the white dashed boxes in
Fig.~\ref{VresFORS1WFI} NE and NW of HCC~007 and carry out the
photometry with the {\it POLYPHOT} task.  {The polygons defined
  for the HCC~007 tails are at average distance of
  $120''\,\mbox{and}\,173''$, where the surface brightnesses in the
  stream are 24.4 and 25.4 mag arcsec$^{-2}$, and those of the
  ``maximal'' Sersic model are 24.3 and 24.8 mag arcsec$^{-2}$. } The
apparent V band magnitudes in the two polygons are $m_v = 14.9$ and
$16.3 $ for the NE, NW sides, respectively. At a distance of $50$ Mpc,
the combined luminosity in the two polygons on the HCC 007 tails is
therefore $L_{\rm V, stream HCC~007} = 2.95 \times 10^9 (\pm 5 \times 10^8)
L_\odot $.  For comparison, the V band luminosity of HCC~007 from the
$V_0$ magnitude measured in \cite{Misgeld08}, $V_0 = 14.18\pm0.01$, is
$L_{\rm V, HCC~007}= 4.5 \times 10^9 L_\odot $. Thus the ratio between
the luminosity in the tail polygons and of the S0 galaxy HCC~007 is
$L_{\rm V, stream HCC~007} / L_{HCC~007} = 0.66$.  This is obviously a
lower limit; especially if the NE tail extends into the off-centered
halo of NGC~3311, its luminosity may be substantially larger than
estimated with the NE polygon.  Thus at least 40\% of the original
luminosity of HCC~007 have already been dissolved from the galaxy.

In summary, the extensive photometric analysis of our V band
images provides strong quantitative evidence for an extended envelope
around NGC 3311 off-centered towards the NE, and for luminous streams
around HCC~026 and HCC~007.  We now wish to establish the kinematic
association of these diffuse structures to their galaxies, with the
goal of understanding their origin.

\section{Long slit spectroscopy for the NGC~3311 outer halo 
and for the tidal stream of HCC~026}\label{sec5}

First we investigate whether the presence of the off-centered,
extended envelope around NGC~3311 is associated with any clear
signatures in the kinematics obtained from deep long slit spectra.  In
Section~\ref{hothalokin} we discuss the mean LOS velocity and velocity
dispersion profiles with respect to the center of NGC~3311. We then
describe the link between kinematic and surface brightness features in
order to constrain the dynamics of the inner galaxy and the outer
halo. In Section~\ref{LSdata} we seek spectroscopic evidence for the
presence of the tidal tails emerging from HCC~026.

\subsection{Kinematic signatures of the off-centered envelope 
 around  NGC~3311}\label{hothalokin}

In \cite{vent10b} we already showed that the halo around NGC~3311 for
$R > 15''$ is dynamically hot.  The kinematic data for NGC~3311 show
an extremely rapid rise from galaxy-typical values of $\sigma_{\rm
  LOS}(R<15'')\simeq 170$ \kms\ to velocity dispersions $\sigma_{\rm
  LOS}(50'') > 400$ \kms\ dominated by the cluster potential.  This
study was based on Gemini long slit spectra along P.A.$= 63^\circ$,
and deep VLT/FORS2 long slit spectra centered on the dwarf galaxy
HCC~026 with P.A.=$142^\circ$. More recently, the VLT/FORS1
spectra along P.A.$= 29^\circ$ published by \cite{Richtler11}
confirmed the large outer velocity dispersion values and the asymmetry
of the $V_{\rm LOS}$ profiles, but also showed deviations between
measurements at different P.A.

The mean LOS velocity $V_{\rm LOS}(R)$ profiles along both P.A.s
increase outwards from NGC 3311 by $\sim 100$ \kms\ to the NE, while
they remain close to the galaxy systemic velocity $V_{\rm NGC3311}(R)=
3800$ \kms\ to the SW.  Thus \cite{vent10b} concluded that the NE outer
halo has slightly more red-shifted mean LOS velocities, $V_{\rm
  halo}\simeq 3900$ \kms, and is offset by $\sim +100$ \kms\ relative
to the NGC~3311 center. The photometric evidence for an offset halo
component to the NE of NGC 3311 shown in Fig.~\ref{VresFORS1WFI}
suggests that the stars in the offset halo could be responsible for
the shift in LOS velocities.

\begin{figure}\centering
\includegraphics[width=8.5cm]{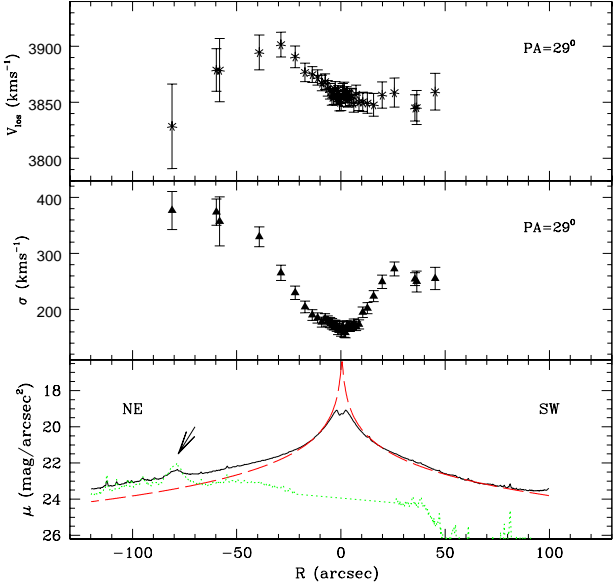}
\caption{Asymmetric kinematics and light distribution of the
  off-centered envelope around NGC 3311. Lower panel: the full black
  line shows the extracted surface brightness profile from the
  VLT/FORS1 V band image along P.A.$=32^\circ$, the red dashed line
  shows the Sersic profile ($n=10.5, R_e=850''$) of the MSM described
  in Sect.~\ref{1dplus2d}, and the green dashed line shows the
  contribution from the off-centered outer envelope along this
  P.A. The arrow indicates the dwarf galaxy HCC~027. Middle and upper
  panels: LOS velocity dispersion ($\sigma_{\rm LOS}$, triangles) and
  velocity measurements ($V_{\rm LOS}$, asterisks) along
  P.A.$=29^\circ$ from \cite{Richtler11} (P.A.$=29^\circ$ shown as
  $R<0$). }
\label{iclkin}
\end{figure}

Building on the photometric model derived in Section~\ref{1dplus2d},
we can check whether either $\sigma_{\rm LOS}(R)$ and/or $V_{\rm
  LOS}(R)$ increase at positions where the surface brightness of the
extended offset envelope is comparable to or brighter than the surface
brightness of the MSM for NGC~3311. In Fig.~\ref{iclkin} we illustrate
this analysis, comparing kinematic and surface brightness profiles
along the galaxy's approximate major axis (P.A.$= 29^\circ$, and
P.A.$= 32^\circ$, respectively). Indeed, both $V_{\rm LOS}$ and
$\sigma_{\rm LOS}$ increase as the surface brightness of the offset
envelope increases from $R\simeq 15''$ outwards.  In particular, the
$V_{\rm LOS}$ values are more red-shifted on the NE side of the galaxy
center than either at the center or in the SW region. However, this
asymmetry peaks at $R\simeq 30''$ and does not closely follow the
estimated relative surface brightness contribution of the off-centred
envelope.

The outermost measurement from \citep{vent10b} of $V_{\rm LOS}=3835$
\kms\ and $\sigma_{\rm LOS} = 429$ \kms\ at $R=100''$ is relevant for
this discussion. These values are measured for a 1D spectrum extracted
from a region in the NGC~3311 halo that is only weakly affected by the
off-centred envelope; the orange-dashed line in
Figure~\ref{DWs_excess} shows the location of this slit region on the
image. These measurements therefore suggest that the symmetric halo is
{\it hot} and {\it approximately at the galaxy systemic velocity}, and
that the off-centred envelope is responsible for
shifting the velocities measured in the NE quadrant of the NGC~3311
halo to more red-shifted values.  We will return to this issue in
Section~\ref{ICPNdwarfs}.

\subsection{Kinematics signatures of the tidal stream around 
HCC~026}\label{LSdata}

We wish now to confirm that the stream-like substructure in the
diffuse envelope NE of NGC~3311 is physically associated with the
dwarf galaxy HCC~026.  The goal is to see whether we can detect
absorption line features in the spectra at the position of the tidal
stream, in addition to the absorption lines from the stars in this
part of the dynamically hot halo of NGC~3311.

\begin{figure} \centering
\includegraphics[width=8.5cm]{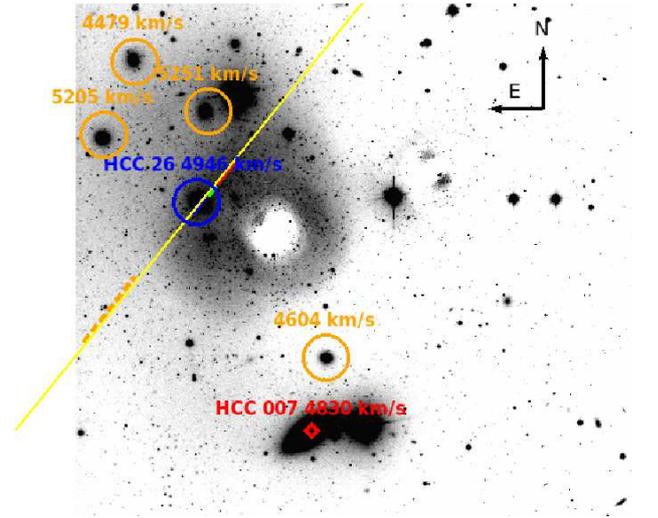}
\caption{Location of dwarf galaxies and long slit on the residual
  VLT/FORS1 V-band image from Fig.~\ref{VresFORS1WFI}.  Here the
  orange circles indicates the dwarf galaxies within 100 kpc of the
  NGC~3311 center; their velocities are reported on the image. The
  {blue} circle shows the dwarf galaxy HCC~026, and the red
  diamond indicates HCC~007.  The object to the right of HCC~007 is a
  foreground star. The yellow line centered on HCC~026 at
  $\mbox{P.A.}=142^{\circ}$ illustrates the position and orientation
  of the long slit used for the kinematic measurements in
  \cite{vent10b} and in this paper.  The orange-dashed section of the
  slit identifies the area where the $V_{\rm LOS}, \sigma$
  measurements at $R=100''$ were taken \citep{vent10b}; see
  Sect.~\ref{hothalokin}.  The red and green parts of the line depict
  the slit sections used to measure the mean velocity of the tidal
  stream.  The FoV is $5'.1 \times 2'.8$; North is up and East to the
  left.}
\label{DWs_excess}
\end{figure}

\begin{figure}\centering
\includegraphics[width=7.5cm]{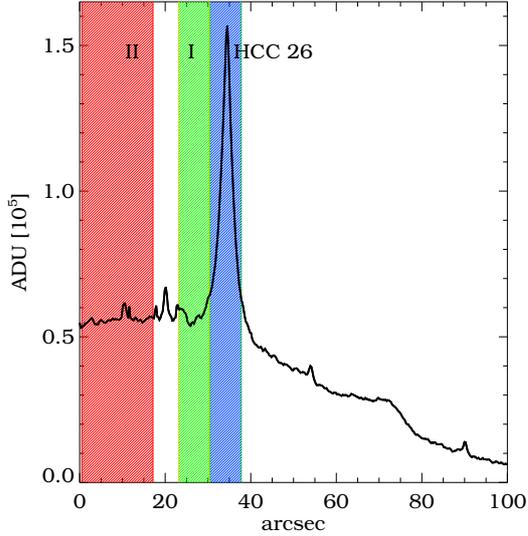}
\caption{Light profile along the slit centered on HCC~026, with the
  different color-shaded areas indicating the slit sections used for
  the extraction of the 1D spectra of HCC~026 (blue) and the tidal
  stream (green and red). Left is SE of HCC~026, right is NW.}
\label{slitsections}
\end{figure}

We use the deep long slit spectroscopic observations of the NGC~3311
halo carried out by \cite{vent10b} with FORS2 on VLT-UT1.  These
spectra extend over the wavelength range $4655\AA\ - 5955 \AA$ and
include absorption lines from H$_{\beta}$, MgI ($\lambda\lambda 5167,
5173, 5184 \AA$) and Fe~I ($\lambda\lambda 5270, 5328 \AA$). They were
acquired with a $6'.8$ arcmin long-slit of width $1''.6$, and
GRISM~1400+V, giving an instrumental dispersion of $0.64\; \AA$
pixel$^{-1}$ and a spectral resolution of $\sigma=90$ \kms.  The long
slit was centered on the dwarf galaxy HCC~026 at
$\alpha=10\mbox{h}36\mbox{m}45.85\mbox{s}$ and $\delta=
-27\mbox{d}31\mbox{m}24.2\mbox{s}$ (J2000), with a position angle
P.A.=$142^{\circ}$ so that is passes through the photometric stream.
HCC~026 itself is seen in projection onto the NGC~3311 halo. The slit
geometry is shown in Figure~\ref{DWs_excess}. Eight exposures of 1800
sec each were taken (in total 4 hrs). In addition to the deep spectra,
the standard G8III star HD102070 and the spectrophotometric standard
star EG~274 were also observed with the same set-up.  The data
reduction followed standard procedures as described in
\citet{vent10b}. The final deep 2D spectrum was then used to construct
radial profiles of the velocity dispersion and mean LOS velocity for
the halo of NGC~3311 from measurements at various positions along the
slit.

Here we want to additionally detect in the spectrum the absorption
lines of the stream superposed onto the main absorption features of
the hot NGC 3311 halo.  We expect these secondary absorption lines to
have strengths of about $15\%$ of those for the NGC~3311 halo, on the
basis of the photometry carried out in Sect.~\ref{1dplus2d}; see
Figs.~\ref{1DVprofs}, \ref{VresFORS1WFI}, and \ref{iclkin}. If the
sub-component projected on the outer halo is a tidal stream emerging
from HCC~026, we expect its absorption lines to occur at significantly
different recession velocity from those of the NGC~3311 diffuse halo
($V_{\rm halo}=3921$ \kms). This is based on the fact that all
of the dwarf galaxies seen at this location, including HCC~026, have
recession velocities of order $5000$ \kms, see Table~\ref{table1} and
\cite{vent11}.

\begin{figure}\centering
\includegraphics[width=8.5cm]{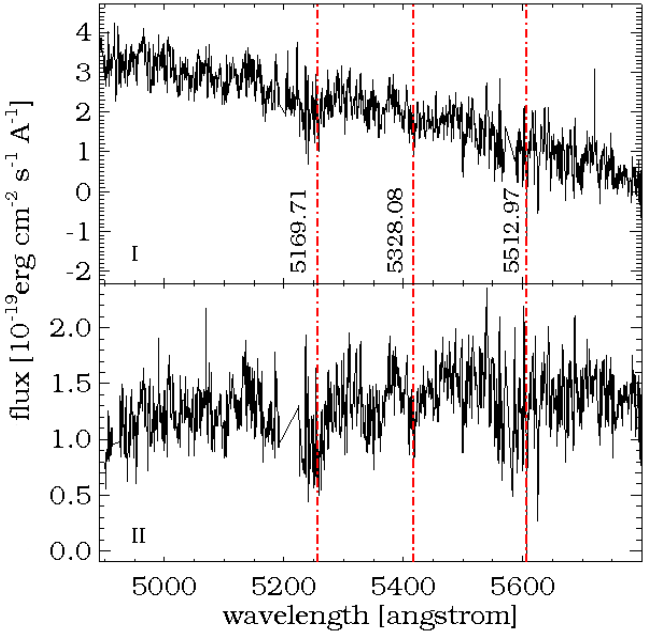}
\caption{ 1D wavelength and flux calibrated residual spectra {\rm I}
  (upper panel) and {\rm II} (lower panel), extracted at the position
  of the light excess. These 1D spectra are obtained by subtracting a
  template model of the NGC~3311 halo from the extracted 1D
  spectra. The vertical dash-dotted lines indicate the locations of
  the main absorption features whose rest frame wavelengths are also
  reported on the figure.}
\label{res}
\end{figure}

We first extract the light profile along the slit and identify those
regions where the continuum is bright enough to provide suitable S/N
for the kinematic measurements. We can identify only two such
regions to the North-West of HCC~026: region {\rm I} - $7''$ wide (30
pixels)- and region {\rm II}- $18''$ wide (75 pixels).  These are
indicated by the green and red slit sections in
Figures~\ref{DWs_excess} and ~\ref{slitsections}. All the flux in each
of these regions is co-added to reach $ S/N \simeq 20$ in the
continuum for the final one-dimensional (1D) extracted spectra. We
must be aware that co-adding all the signal from an extended portion
of the slit causes a broadening of the absorption lines, but this does
not affect our goal, which is the detection of a secondary absorption
line component at a different LOS velocity from that associated with
the hot halo component.

\begin{figure*}\centering
\includegraphics[width=19.5cm]{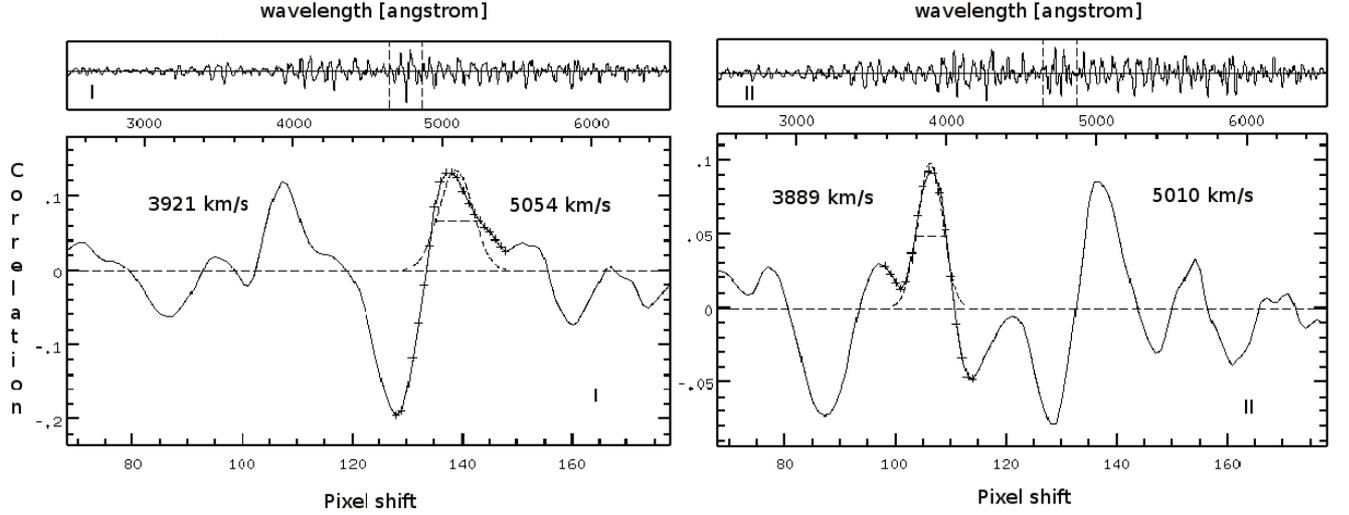}
\caption{Cross correlation functions used for kinematic measurements
  in two slit sections. \textit{Upper panels}: Fourier
  cross-correlation functions computed between the residual (excess)
  spectra and the spectrum of the G dwarf star HD102070 for spectrum
  {\rm I} (left) and {\rm II} (right).  \textit{Lower panels}:
  Enlarged view of the Fourier cross-correlation function peaks. The
  double peaks in both Fourier cross-correlation functions indicate
  the presence of two sets of absorption line features at different
  velocities in the residual spectra. The more red-shifted peaks in
  the Fourier cross-correlation are at velocities $V_{\rm LOS,1} =
  5054,\, 5010\, (\pm 55)$ \kms, and the secondary bluer peaks are at
  $V_{\rm LOS,2} = 3921,\, 3889\, (\pm 55)$ \kms.}
\label{correlation}
\end{figure*}

To measure the stellar kinematics, we restrict ourselves to the
central wavelength range $4800\AA\ < \lambda < 5800 \AA$ of the
extracted 1D spectrum, and use the ``penalized pixel-fitting'' method
\citep[PPXF][]{Cappellari04}. In the PPXF method, stellar template
stars from the MILES library \citep{Sanchez07} are combined to fit the
1D spectrum while simultaneously determining the mean LOS velocity and
velocity dispersion from the absorption lines.  From the best PPFX fit
to the co-added spectrum we obtain values for these kinematic
parameters which are consistent with the halo kinematics measured by
\citet{vent10b}, as well as a best-fit stellar template for the halo of
NGC~3311.

From the photometry carried out in Section~\ref{1dplus2d}, the
luminosity from the tidal stream amounts to $\sim 15\%$ of the
total light in the region sampled by this spectrum.  Attempting a
direct spectral decomposition of the stellar halo of NGC 3311 and the
light excess as in \citet{Coccato11a} was unsuccessful. To be able to
detect the weak kinematic signal from the excess of light, we
therefore need to subtract the main halo contribution.  We achieve
this by taking the best-fit stellar template spectrum obtained by PPFX
for the total extracted science spectrum, multiply it by a suitable
fraction ($0.85$) and subtract it off the extracted spectrum. This
procedure when applied to the 1D spectra {\rm I} and {\rm II} leaves
the residual spectra shown in Figure~\ref{res}.
 
Clearly, the S/N of the residual spectra is not high enough for a
direct pixel fitting, although the main absorption features are
readily identified, and we must now use a different approach. We use
the {\it RV.FXCOR} task in IRAF to identify the velocity components in
the residual spectrum; this task implements the Fourier
cross-correlation technique of \cite{fxcor79}.  The upper panels of
Figure~\ref{correlation} show the Fourier cross correlation functions
computed with {\it RV.FXCOR} between the residual 1D spectra {\rm I}
and {\rm II} and the extracted 1D spectrum of the star HD102070. All
spectra have their continuum fitted and subtracted-off, and only the
wavelength interval $4800\AA\ < \lambda < 5800 \AA$ is used.

In the lower panels of Fig.~\ref{correlation}, the region centered
around the two strongest correlation peaks is enlarged. This reveals
the presence of two components at different velocities in both
spectra: the strongest peak is found at $5054$ \kms, $5010$ \kms\ and a
second peak is detected at $3931$ \kms, $3889$ \kms, for spectrum {\rm
  I} and {\rm II} respectively. The error in each velocity measurement
is $\pm 55$ \kms.  { The estimated rms noise in the velocity range
  2000 - 8000 \kms in the two lower panels of Fig.~\ref{correlation}
  is $\simeq 0.066$ and $\simeq 0.038$. The height of both peaks in
  the cross correlation spectrum in each panel is therefore $\sim 2$
  times the amplitude of the noise; because the two panels are for two
  independent spectra} we therefore judge these signals to be
significant.  The bluer velocity peaks correspond to a component from
the extended halo light that is still present in the residual spectra
at an average LOS velocity of $3905$ \kms, see also
Sect.~\ref{hothalokin}. The peaks at more redshifted LOS velocities
indeed provide evidence for a second component at $\Delta V \simeq
+1200$ \kms\ relative velocity with respect to the systemic velocity
of NGC~3311 ($V_{\rm NGC3311} = 3800$ \kms), and $\Delta V \simeq
+1100$ \kms\ relative to the halo ($V_{\rm halo}=3921$ \kms). The
average LOS velocity obtained by combining the measurements from
spectra {\rm I} and {\rm II} is $5032 \pm 38$ \kms. For comparison,
the LOS velocity of HCC~026 is $4946 \pm 4$ \kms\ \citep{vent10b}.

In summary, the stars in the stream NE of HCC~026 have a LOS velocity
very similar to that of HCC~026 itself but $\Delta V \simeq +1200$
\kms\ different from the systemic velocity of NGC 3311. This is a
necessary condition for the stream around HCC~026 being a tidal
stream, but could also be consistent with the stream stars being
part of a shell which was stripped from a larger galaxy together with
HCC~026. However, combining our kinematic result with the result of
\citet{Coccato11b} that the stellar population at the NW stream
position is more metal poor than in the surrounding outer halo,
consistent with the superposed stream stars having similar metallicity
as the dwarf galaxy HCC~026 itself, makes a strong case that the stars
in the stream were indeed tidally stripped from HCC~026.

\section{Correspondence between surface brightness 
 components and kinematical structures measured 
 with planetary nebulae, and the velocities along 
 the HCC 007 tail }\label{ICPNdwarfs}

To begin with, we summarize the results from observations of PN
kinematics in the Hydra I cluster core, in order to see how they may
help in understanding the physical structures identified in the
photometry.  In \cite{vent11}, we measured LOS velocities for 56 PNs
in a $6.8' \times 6'.8$ field covering the central $(100 {\rm kpc})^2$
of the Hydra~I cluster, using multi-slit imaging spectroscopy
\citep[MSIS, ][]{Gerhard05,Arnaboldi07}.  We identified different
velocity components in the PN LOSVD and described their spatial
distributions. In brief, we found:
\begin{itemize}
\item A broad, asymmetric central velocity component in the PNs LOSVD,
  peaked at $\sim 3100$ \kms and with $\sigma \simeq 500$ \kms.  The
  spatial distribution of these PNs has significant substructure,
  depending on the LOS velocity. It is not clear whether this
  component traces the symmetric halo around NGC~3311.
\item A minor blue-shifted velocity component, centered at $\sim 1800$
  \kms, whose PNs have an elongated distribution along the North/South
  direction.  There is no galaxy with similar velocity inside the
  $6.8' \times 6'.8$ field centered on NGC~3311, but two such galaxies
  are located near the perimeter of the field. The association of these
  with the blue-shifted PNs is possible but unclear, however.
\item A narrow red-shifted velocity component, around $V_{\rm
    LOS}\simeq 5000$ \kms. These red-shifted PNs show a concentration
  in the NE quadrant of NGC~3311, and are correlated spatially and in
  LOS velocity with a group of dwarf galaxies, which are observed
  within $(100$ kpc)$^2$ of NGC~3311.
\end{itemize}

\begin{figure}\centering
\includegraphics[width=9.0cm]{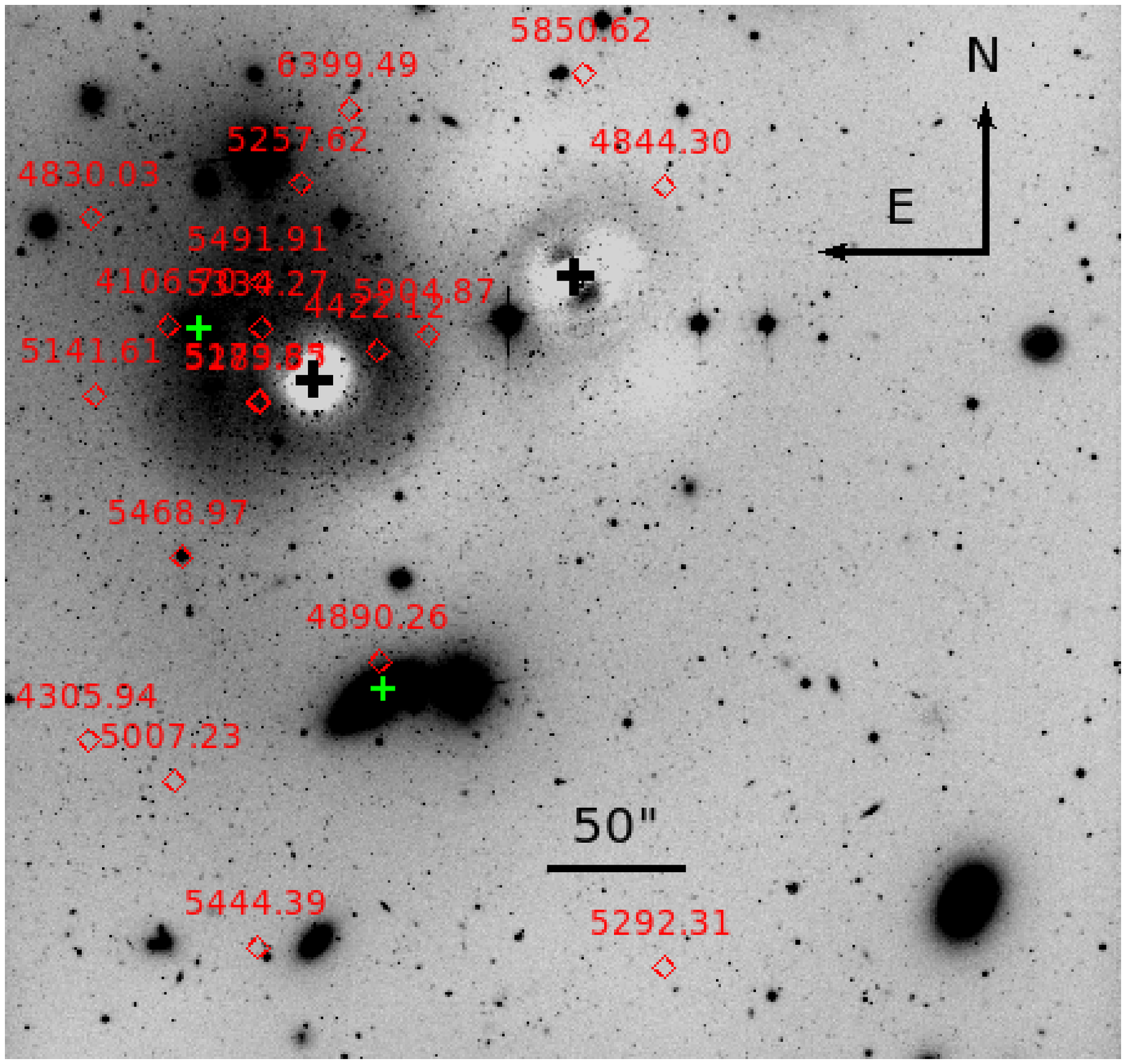}
\caption{Positions and velocities of PNs (red diamonds, labels
  indicate the PNs' $V_{\rm LOS}$ in \kms) associated with the red
  peak at 5000 \kms\ in the PN LOSVD of the Hydra~I cluster, from
  \cite{vent11}. The red peak PNs are superposed on the residual
  V-band image, which shows the substructures in the diffuse light in
  the Hydra~I cluster core.  The black crosses indicate the position
  of NGC~3311 (center) and NGC~3309 (upper right), respectively. The
  green crosses indicate HCC~026 and HCC~007. {The FoV is
    $6'.8\times 6'.4$. }}
\label{red_PN}
\end{figure}

In Figure~\ref{red_PN}, we plot the PNs associated with the
red-shifted peak of the PN LOSVD \citep{vent11} on the residual
VLT/FORS1 V band image with respect to the MSM for NGC~3311.  This
shows that a major fraction of these red-shifted PNs are concentrated
in the NE quadrant where the offset halo is found. We count 10 PNs
superposed on the offset envelope. Their average LOS velocity is
$V_{\rm redPNs, LOS} = 5095$ \kms\ and $\sigma_{\rm redPNs, LOS} =
521$ \kms. The average LOS velocity is similar to that of HCC 026
($V_{\rm HCC 026} = 4946$ \kms) and to the value for the second
component measured in Sect.~\ref{LSdata} ($5032$ \kms\ average).  With
a relative offset of $1200$ \kms\ from NGC 3311 and a contribution of
$\simeq 15\%$ to the surface brightness, the shift in the predicted
mean velocity would be of order $180$ \kms, consistent with the
measured value. However, we found in Sect.~\ref{Ltails} that the
off-centered envelope contributes $\sim30\%$ of the surface brightness
in the NE region, and this value is consistent with the value
indicated at the HCC~026 position in
Figs.~\ref{1DVprofs},~\ref{iclkin}. Thus, while from their
spatial location the red-shifted PNs appear to trace the off-centered
envelope, it is not clear whether they trace all the stars in it.

Therefore, from the off-centered envelope's luminosity estimated in
Sect.~\ref{Ltails}, $L_{\rm V,NE,env}= 1.2 \times 10^{10}\, (\pm 6.0
\times 10^8)\,L_\odot$, we can derive an upper limit for the
luminosity specific PN number $\alpha$ for this stellar population.
The measured value is $\alpha = 1/ 1.2 \times 10^9$ PN $L_\odot^{-1}$;
once we correct this number for the limiting magnitude of these MSIS
observations $\alpha_{\rm TOT} = 82 \times \alpha$ (see \cite{vent11}
for further details), we obtain $\log (\alpha_{\rm TOT}) =
-7.18$. This value for $\alpha_{\rm TOT}$ is similar to that for old
stellar populations which are observed in the M31 bulge and S0s
galaxies \citep{Buzzoni06}.

For the symmetric halo of NGC~3311 the $\alpha$ parameter must be
lower. Within $25'' < R < 120''$, we count 12 PNs from the central
velocity component \citep{vent11}. If these trace the total luminosity
of the symmetric halo, $L_{\rm V,halo} = 7.5 \times 10^{10}\, (\pm 2.0
\times 10^9)\,L_\odot$, the measured $\alpha = 1 / 6.2 \times 10^9$ PN
$L_\odot^{-1}$ and the corresponding $\log (\alpha_{\rm TOT})$ is
$-7.87$, a factor 5 lower than the value for the stellar population in
the offset envelope. In fact, the LOSVD of the central PN component
is quite axisymmetric, and very few PNs at the systemic velocity
of NGC 3311 are found on the inner symmetric halo. Thus the
$\log (\alpha_{\rm TOT})$  for this component may be even lower.

Superposed on the off-centered envelope, there are two tidal streams
in the Hydra~I core; they are shown in Figs.~\ref{v2dmod} and
\ref{VresFORS1WFI}. In Sect.~\ref{LSdata} the direct measurement of
$V_{\rm LOS}$ of the stream around HCC~026 is $5032 \pm 38$ \kms, and
the systemic velocities of HCC~026 and HCC~007 are $V_{\rm HCC~026} =
4946 $ \kms, $V_{\rm HCC~007} = 4830 $ \kms\ respectively. Four of
the PNs plotted in Figure~\ref{red_PN} are consistent with being
located on the NE HCC 007 stream (one is on the polygon used to
estimate the luminosity). These indicate a systematic velocity
gradient along the stream, increasing from $4830$ \kms\ east of the
off-centered envelope to a maximum of $5470$ \kms\ near the closest
approach to NGC 3311, and decreasing again to $4890$ \kms\ near HCC
007 itself.

Assuming that the stellar population in these streams is similar to
that in the off-set envelope, we would expect $2\pm 1$ PN in the
polygon on the NE tail of HCC~007, and no PNs on the SE tail or on the
stream around HCC~026.  In Figure~\ref{red_PN}, we count one PN
associated with the corresponding part of the HCC 007 stream, which is
consistent with the prediction.  There is not enough light in the
stream around HCC~026 to expect any PN to be detected within the
limiting magnitude of the \cite{vent11} survey.

\section{Discussion}\label{discussion}

\subsection{The peculiar outer halo of NGC 3311}\label{discusshalo}

{\sl Outer halo or intracluster light? --} Based on V-band photometry
out to $\sim 100''$ ($\sim 25$ kpc), the outer halo of NGC~3311 can be
represented by a symmetric Sersic model with large $n \simeq 10$, and
an additional off-centered component with centroid shifted by about
$50''$ to the NE (Sect.~\ref{symmodels}).  The off-centered component
can be described by a (flatter) exponential profile. However, it is
not the extended light profile which distinguishes this central
cluster galaxy from other luminous elliptical galaxies, but its very
steeply rising velocity dispersion profile (VDP) $\sigma(R)$. From a
central value of $\sim 170$ \kms, the VDP rises to $\sigma\simeq 230$
\kms\ at $R=15''\simeq 3.7$ kpc, and then on to $\sigma=300-450$ \kms\
at $R=50''\simeq 12$ kpc \citep{vent10b,Richtler11}. The steep rise of
the velocity dispersion profile corresponds to a steep increase in the
enclosed mass. Within $R=20$ kpc, the total dark matter mass inferred
from X-ray observations is $\sim 10^{12} M_\odot$
\citep{Hayakawa04}. Thus NGC~3311 is located at the center of the dark
matter cusp of the Hydra I cluster.

In the NW region dominated by the off-centered halo component, the
velocity dispersion is particularly high ($\sim 450$ \kms), about
$60\%$ of the galaxy velocity dispersion in the cluster core
\citep{vent10b}. Its mean LOS velocity is also shifted by $\sim 100$
\kms\ with respect to NGC~3311. Many PNs located on the off-centered
envelope move at LOS velocities of $+1200$ \kms\ with respect to the
NGC~3311 center \citep{vent11}, but also the LOSVD of the 'central PN
component' of \citet{vent11} is quite asymmetric. The off-centered
envelope may therefore equally well be considered as part of the ICL
in the cluster core; because of the steeply rising VDP, distinguishing
outer halo from ICL is difficult \citep[see][]{Dolag10}.

{\sl Dynamically hot outer halos in other BCGs? --}
A galaxy with a dynamically hot outer halo similar to NGC 3311 is
NGC~6166, the cD galaxy in the cluster A2199 \citep{Kelson02}.  In
this system, the velocity dispersion first decreases from the central
value of $300$ \kms\ to $200$ \kms\ within a few kiloparsecs, and then
steadily rises to $660$ \kms\ at a radius of 60 kpc, nearly reaching
the velocity dispersion of the cluster ($\sigma_{A2199}=775\pm50$
\kms). This is similar but less extreme than the case of NGC~3311
studied here, which does not show the central decrease.  However, such
dynamically hot outer halos in BCGs are rare: e.g., for NGC~1399 in
the Fornax cluster \citep{emily10}, NGC~4874 and NGC~4989 in Coma
\citep{coccato101} the outer VDPs are flat, while for M87 in the Virgo
cluster the VDP first rises out to $250''$ and then falls steeply
\citep{Doherty09,Murphy11}. 

{\sl Asymmetry in X-ray observations of the Hydra I cluster core. --}
Independent evidence for asymmetries in the Hydra I cluster center
comes from Chandra and XMM observations
\citep{Hayakawa04,Hayakawa06}. These authors report about an extended
emission $\sim 1'.5$ in the direction NE from NGC~3311 relative to the
iso-surface brightness contours further out, with an angular scale of
about $1.5'$, see Fig.~2 in \cite{Hayakawa04} and Fig.~5 in
\cite{Hayakawa06}. The X-ray surface brightness maps show a morphology
reminiscent of the off-centered halo in the stellar light
distribution.  The extended X-ray emission NE of NGC~3311 is also the
brightest among several high metallicity regions in the Hydra~I
cluster, and its associated gas mass is larger than that in the
compact X-ray halo of NGC~3311.  \cite{Hayakawa04}
interpret this high metallicity region $1'.5$ to the NE of NGC~3311 as  
the result of gas stripping of NGC~3311, which would imply a relative
motion of NGC~3311 with respect to the surrounding outer halo. It may
be possible that the collision with the dark matter halo of the group
including HCC~007 could have induced such a relative motion; this
requires further study.

{\sl Off-centered outer envelopes in BCGs. --}  
Separate outer components in BCGs are frequent; the work of
\cite{Gonz05} for a sample of 24 clusters shows that a two component
fit to the light profiles of BCGs provides an improved match to the
data, and that the two photometric components are misaligned in $60\%$
of the sample. \cite{Gonz05} reported also the case of Abell 1651,
where the two components have different centers, with the outer
component off-centered by about 15 kpc in linear distance. This is
similar to the separation between the center of the off-centered
envelope described in Sect.~\ref{symmodels}, and also the center of the
extended X-ray emission, from the inner parts of NGC~3311.

{\sl Stellar population in the outer halo of NGC~3311. --}
\cite{Coccato11b} measured the line strength indices for the stellar
population in the NGC~3311 halo along a $6'.8$ slit centred on HCC~026
and with P.A.=$142^\circ$. These measurements show the effect of the
stars in the HCC~026 tail: in this region, the metallicity is lower,
[Fe/H]$ = -0.73 \pm 0.06$, compared to the average value for the outer
halo away from the tail, where [Fe/H]$_{\rm halo} -0.34 \pm 0.05$.
The metallicity in the tail region is consistent with a $\sim 20\%$
contribution of stars with the metallicity of HCC~026. Unfortunately,
in the halo region away from the HCC~026 tail, these data do not have
sufficient S/N to test whether the stars in the offset envelope are a
different stellar population from those in the symmetric halo.

\subsection{Tidal tails and streams around NGC 3311}
\label{discusstails}

{\sl Comparison of tails in Hydra I with those around other galaxies. --}
It is relevant to compare the luminosities of the streams in the
Hydra~I core with those observed in the halos of bright ellipticals in
the Virgo cluster by \cite{janow+10} . The HCC~026 stream identified
here with total luminosity slightly below $10^9 L_{\odot,V}$ is
significantly brighter than any of the Virgo streams, and the HCC~007
stream is several times more luminous still. Also, the tidal streams
seen around the Virgo cluster ellipticals have much fainter peak
surface brightness (brightest: $\mu_V = 25.7$ mag~arcsec$^{-2}$;
average: $\mu_V = 27.6$ mag~arcsec$^{-2}$) than those measured in the
Hydra~I streams, which are $\mu_V = 24.8$ mag~arcsec$^{-2}$ for the
HCC~026 stream and $\mu_V = 24.4$ mag~arcsec$^{-2}$ for the HCC~007
stream. 

Streams with as low surface brightness as those observed in Virgo
would not be detectable in our Hydra images, thus could easily be
abundant in this region. { In fact, the two small objects approximately
near the middle of the polygon on the NE stream of HCC~007 in
Fig.~\ref{VresFORS1WFI} may be connected by a small stream within
the HCC~007 tail, as suggested especially by the right panel of
Fig.~\ref{v2dmod}. This small stream, if confirmed, would have a
luminosity more similar to the Virgo streams, albeit a higher surface
brightness}.

\begin{table*}
\begin{center}
\begin{tabular}{l*{5}{c}r}
\hline\hline 
Galaxy&$\alpha$(2000)&$\delta$(2000)& $M_V$  & v     \\
       &  [h:m:s]    &[$\circ$:':'']& [mag] &  [\kms] \\
\hline
HCC 019 & 10:36:52.573  & -27:32:16.34 & 16.91 & 5735$\pm$55 \\
HCC 022 & 10:36:40.373  & -27:32:57.68 & 18.23 & 4605$\pm$37 \\
HCC 023 & 10:36:48.911  & -27:30:01.49 & 18.07 & 4479$\pm$44 \\
HCC 024 & 10:36:50.140  & -27:30:46.20 & 17.75 & 5270$\pm$32 \\
HCC 026 & 10:36:46.000  & -27:31:25.10 & 17.87 & 4946$\pm$04$^*$\\
HCC 027 & 10:36:45.700  & -27:30:31.30 & 18.48 & 5251$\pm$89 \\
HCC 007 & 10:36:41.200  & -27:33:39.60 & 14.18 & 4830$\pm$13\\
\hline
\end{tabular}
\end{center}
  \caption{Dwarf galaxies in the NGC 3311 field. Magnitudes and
    line-of-sight velocities are from \cite{Misgeld08}. $(^*)$ 
    value from Ventimiglia (2011), Ph.D. Thesis. }
\label{table1}
\end{table*}

We can expand this comparison to the streams around bright local
elliptical galaxies analyzed by \cite{tal+09}.  The Hydra~I streams
are detected at radii larger than $\sim 23$ kpc from the center of
NGC~3311, and the HCC~007 tail reaches at least $110$ kpc in total
projected length, see the measurement in Sect.~\ref{2dgalfitv}. These
numbers for the Hydra~I streams are consistent with the average radius
of occurrence of the streams studied by \cite{tal+09}, and the linear
extent of some of them. Unfortunately, \cite{tal+09} do not give
luminosities for their streams.

{\sl The streams are tidal tails. --} The streams emerging from
HCC~026 and HCC~007 can be interpreted as the result of tidal
disruption. In the case of HCC~007, the sheer size of the stream (the
two tails together span $\sim 110$ kpc) is difficult to explain by any
other model. For HCC~026, the similar LOS velocity of the stream with
the galaxy HCC~026 could still allow a model in which the stream stars
are part of a shell which was stripped from a larger galaxy together
with HCC~026. However, \citet{Coccato11b} showed that the metallicity
obtained with a single stellar population model in the stream region
is lower than in the diffuse stellar halo of NGC~3311 away from the
stream, and that this lower metallicity value is consistent with a
composite population of stars from HCC~026 and stars from the stellar
halo of NGC~3311 in approximately the proportion implied by the
surface brightnesses of both components.  This makes a strong case
that the stars in the HCC~026 stream were indeed tidally stripped from
HCC~026. It also shows that a fruitful way to investigate the origin
of the diffuse light features in Hydra I and elsewhere is to measure
stellar population parameters from deep spectroscopy.

Given their redshifted velocities, both the HCC~007 and HCC~026
streams would likely be located now on the distant side of the Hydra~I
cluster core, with their stars being on slightly different orbits than
the galaxies they were once bound to.

{\sl Comparison with simulated tails. --}
The overall morphology of the two streams in the Hydra~I cluster core
is similar to those of tidal streams in the simulation investigated by
\citet{Rudick09} to study the formation of intracluster light. In
particular, the tidal streams in their case G1 reproduce one
characteristic property of the streams discovered here, especially for
HCC~007, that one side of the stream is significantly brighter than
the other. In the simulation of \citet{Rudick09}, the $\sim 200$ kpc
long tidal arms in G1 formed as a consequence of the interaction of a
disk galaxy with the cluster cD, with a pericenter distance of $\sim
100$ kpc. Most of the stars in the tidal streams in this system were
unbound from the disk galaxy shortly after the pericentre passage, and
once formed, the streams decay in $\sim 1.5$ orbital times.  The fact
that tidal streams are visible around HCC~026 and HCC~007 thus
indicates that both galaxies have recently passed the pericenters of
their orbits.

\subsection{A group of galaxies in disruption in the Hydra I cluster core}

{\sl All the galaxies in disruption. --} As already reported by
\cite{vent11}, there are no galaxies in the Hydra I core at
cluster-centric radii ($ < 100$ kpc) with velocities around the
cluster systemic velocity.  However, several dwarf galaxies at high
velocities $V_{\rm LOS} \ge 4500$ \kms and the S0 galaxy HCC~007 are
located in this region.  In Table~\ref{table1} we list the sky
coordinates, apparent total V-band magnitudes, and LOS velocities for
these galaxies from \cite{Misgeld08}.  These galaxies are part of a
well-defined cluster substructure both in velocity and spatial
distribution, as already commented by \cite{vent11} and shown in
Figure~\ref{DWs_excess}. For two of these galaxies, HCC~026 and
HCC~007 we have been able to detect tidal tails, showing that they are
being disrupted by the tidal field near the cluster center. For the
others, the tidal effects may not yet be strong enough to have lead to
detectable tidal tails, or their tails have dispersed below the
detection limit. Independent of this, the observed tidal streams are
the consequence of the recent infall of a cluster substructure
containing the galaxies at $V_{\rm LOS} > 4500$ \kms.

{\sl Where have the central galaxies gone? --}
We now return to the observed absence of galaxies at cluster-systemic
velocity in the central 100 kpc around NGC~3311 (a similar result was
found in the NGC~5044 group, \citet{Mendel09}).  A plausible
explanation is that such galaxies are no longer seen in the central
region of the cluster because they were all disrupted in the past
during close encounters with the central galaxy NGC~3311 and with the
dark matter cusp at the cluster center \citep{Afalt05}. In this case,
their former stars would now be part of the diffuse stellar component
in the Hydra~I core, including the halo around NGC~3311.  It is
interesting to ask whether most of the stars in the outer halo could
have originated from disruption of small galaxies in the way that
seems to be currently on-going, or whether the majority of the halo
stars come through a different channel, the tidal disruption of the
halos of massive elliptical galaxies prior to merging with the cluster
cD \citep{Murante07,Puchwein10}. Again, analysis of the stellar
population properties of dwarf galaxies and of the outer halo stars is
important for answering this question \citep{Coccato10,Coccato11b}.

{\sl Phase-mixing and formation of ICL from the tidally dissolved
  stars. --} Assuming that the interaction with NGC~3311 and the
Hydra~I cluster core ultimately disrupts the dwarf galaxies seen close
to NGC~3311 completely, all of their stars would phase-mix and
distribute in the central cluster potential. Converting the
uni-directional velocity of their orbits of now $\sim 1200$ \kms
relative to NGC~3311 to $3\sigma^2_{\rm eq}$, the resulting
$\sigma_{\rm eq}\simeq 700$ \kms. This would be somewhat increased by
adding the contribution from the unknown transverse velocities. On the
other hand, some of the kinetic energy of these stars would be
converted to potential energy during the phase-mixing, lowering the
final $\sigma_{\rm eq}$. For comparison, the velocity dispersion of
the NGC~3311 halo near the current position of HCC~026 is $\sigma_{\rm
  halo}\simeq 400$ \kms.  This suggests that the stars from the
galaxies that are presently tidally disrupted will end up at larger
radii than their current position, and with somewhat larger dispersion
than the observed $\sigma_{\rm halo}\simeq 400$ \kms.  I.e., they will
end up in the outermost halo and intracluster light around NGC~3311
and in the Hydra I cluster core.

\section{Summary and conclusions}\label{sumcon}

In this work we extend our investigation of the properties and origin
of the diffuse light in the Hydra~I cluster. Combining surface
photometry with kinematic information from long slit and planetary
nebula (PN) data, we find an off-centered, diffuse outer halo around
the central galaxy NGC 3311, and show that at least two galaxies are
currently disrupted in the cluster core, adding their stars to the
outer halo and intracluster light around NGC~3311. More specifically,
our results are as follows:

Structural parameters are derived for the two giant elliptical
galaxies in the cluster core, NGC 3309 and NGC 3311, using V-band
imaging data obtained at the ESO/MPI 2.2m telescope and archival
VLT/FORS1 V band data. While the light distribution of NGC~3309 is
reproduced by a single Sersic profile, that of NGC~3311 is
characterized by several components.  Outside the nuclear regions
which are affected by a dust lane and by bright luminous knots, the
bright regions within $30''$ follow an $R^{1/4}$ law. In the deep V
band data, these regions together with the symmetric part of the outer
halo can be described by a Sersic law with $n \simeq 10$.

The residual image, obtained after subtracting the two-dimensional
model of the two bright galaxies, shows an additional extended
envelope centered at $\sim80''$ to the North-East of NGC~3311.  This
off-centered envelope is approximately described by a exponential
profile and contains $L_{\rm V,NE,env}= 1.2 \times 10^{10}\, (\pm 6.0
\times 10^8)\,L_\odot$. This corresponds to $\sim50\%$ of the
luminosity of the symmetric halo in the same region, and $\sim15\%$ of
the luminosity of the entire symmetric halo in the radial range
$25''-120''$.

Furthermore, the diffuse light in the Hydra~I core harbors two tidal
streams emerging from the dwarf galaxy HCC~026 and the S0 galaxy
HCC~007 (see Fig.~\ref{VresFORS1WFI}). The total luminosity in the NW
part of the HCC~026 stream is $L_{\rm V,NW,HCC~026} = 4.8 \times 10^8
(\pm 8 \times 10^7)\, L_\odot$, with an average surface brightness
$\mu_V = 24.8 \pm 0.2$ mag arcsec$^{-2}$, $\sim 15\%$ of that in the
extended envelope and symmetric halo at the stream position.  

Analysis of a deep spectrum in this region indicates that the NW
stream has a LOS-velocity similar to HCC~026 itself ($\sim 5000$ \kms;
see Sect.~\ref{sec5}), and that it consists of stars with similar
metallicities as the stars of HCC~026 \citep{Coccato11b}. These results
favour the interpretation that the HCC~026 stream has been tidally
dissolved from HCC~026, over one where both the dwarf galaxy and the
stream were dissolved from a larger galaxy. The luminosity of the
combined NW and SE parts of the stream is several times the current
luminosity of HCC~026, $L_{\rm V,HCC~026} = 1.5 \times 10^8 L_\odot$,
so that this dwarf galaxy has by now been mostly dissolved by the
tidal field. Both HCC~026 and its tidal tails have a relative LOS
velocity of $\sim 1200$ \kms\ with respect to NGC 3311 ($V_{\rm NGC
  3311}\simeq 3800$ \kms).

The second tidal stream around the S0 galaxy HCC~007 extends over at
least $\sim 110$ kpc on both sides of this galaxy. The tail is fairly
thick and is brighter on the NE side where $\mu_V = 24.4 \pm 0.5$ mag
arcsec$^{-2}$. {On the NW side it is about one magnitude fainter
  and less certain.} We measured a luminosity of $L_{\rm V,NE,HCC~007}
= 2.9 \times 10^9 (\pm 5 \times 10^8)\, L_\odot$ in the brightest
region of the NE tail. The NE tail appears to join the outer part of
the off-centered halo east of NGC 3311; if so, its total luminosity
could be significantly higher.  For comparison, the total current
luminosity of HCC 007 is $\sim 4.7 \times 10^9L_{\odot}$. This galaxy
has therefore lost of order 50\% of its stars.

Four PNs are found on the NE HCC~007 tail which allow a preliminary
measurement of its LOS velocity. The largest PN velocity ($5470$ \kms)
is found SE of NGC 3311 in a region dominated by tail stars. One PN is
close to HCC~007 itself and has a velocity within $60$ \kms\ of the
galaxy's systemic velocity.

The morphologies of these streams are similar to those of tidal
streams generated when galaxies are disrupted on highly radial orbits
through the cluster center, shortly after the time of closest approach
with the cluster cD \citep{Rudick09}.

Superposed on the off-centered outer halo of NGC 3311, we also find a
number of PNs from the same sub-component of the PN LOSVD centered at
$5000$ \kms, as well as a group of dwarf galaxies with similar
velocities. This suggests a physical association between the
off-centered halo, the two tidal streams, and an entire group of
galaxies at about $5000$ \kms systemic velocity, including HCC~026,
several other dwarf galaxies, and HCC~007, which are currently falling
through the core of the Hydra I cluster and have already been
partially disrupted.

What fraction of the off-centered halo is due to stars dissolved from
these galaxies is, however, not constrained well by our
measurements. The deep spectra show only a modest shift in mean LOS
velocity in the NE halo, and the photometric, kinematic and stellar
population analysis of the HCC~026 tail suggests that the component at
$5000$ \kms\ LOS velocity contributes only a fraction of the inferred
off-centered halo at this location. Furthermore, a number of PNs with
moderately to highly blue-shifted velocities are also seen superposed
on the outer halo of NGC 3311 \citep{vent11}. To resolve this requires
further kinematics and stellar population analysis with deep
spectroscopy at different positions in the halo.

This work provides a vivid example of how strong tidal forces cause
morphological transformations of galaxies falling through the cores
of galaxy clusters.  It also shows that stars are currently being
added to the diffuse light in the Hydra~I cluster core. The stars in
the streams around HCC~026 and HCC~007 must have been unbound from
their parent galaxies during the on-going close passage through the
high-density center of the cluster, and are now on slightly different
orbits from their once parent galaxies. Ultimately they will phase-mix
in the potential and add to the diffuse outer envelope of NGC~3311 and
to the intracluster light in the cluster core.

The on-going accretion of a group of galaxies will add about several
$10^9 L_{\odot, V}$ to the outer halo and intracluster light around
NGC~3311, about 5\% of the total light in the NGC~3311 symmetric halo.
Tidal dissolution of small galaxies is thus an important channel for
creating diffuse light in clusters. It is unlikely, however, that this
is the main mechanism for the origin of the symmetric halo around
NGC~3311. Both simulations \citep{Murante07,Puchwein10} and studies of
diffuse light in clusters like Coma and the star pile cluster
\citep{Gerhard07, Salinas11} indicate that the disruption and merger
of a giant galaxy is the more efficient route to creating large
amounts of diffuse light.  Deep spectroscopy to measure Lick indices
and stellar abundances in different regions of the offset envelope and
halo of NGC 3311 will be important to address this issue further.

\begin{acknowledgements}
  The authors thank the referee for his/her report, the
  ESO VLT staff for their support during the observations, and Ken
  Freeman for helpful discussions during the course of this project.
  They also thank Ricardo Salinas and Tom Richtler for sending their
  NGC~3311 kinematic data in electronic form. EI acknowledges support
  from ESO during several visits while this work was completed. LC
  acknowledges funding from the European Community's Seventh Framework
  Programme (FP7/2007-2013/) under grant agreement No 229517. This
  research has made use of the 2MASS data archive and the NASA/IPAC
  Extragalactic Database (NED), and of the ESO Science Archive
  Facility.
\end{acknowledgements}

\bibliography{P2Phot}

\end{document}